\documentclass[10pt]{article}
\usepackage{amsmath}
\usepackage{amsfonts}
\usepackage[english]{babel}
\usepackage[ansinew]{inputenc}
\usepackage{graphicx}
\usepackage{graphpap}
\usepackage{accents}
\usepackage{psfrag}

\def\defi{{\stackrel{\mbox{\tiny {\bf def}}}{\,\, = \,\, }}}
\def\eqN{{\stackrel{\mbox{ \tiny $\bm{\N}$\,\,\,}}{=  }}}
\newcounter{remark}
\newcommand{\ep}{\hspace{\stretch{1}}$\Box$}
\newcommand{\vs}{\vspace{5mm}}

\def\gMm{g}
\def\onehalf{{\textstyle \frac{1}{2}}}

\def\partiald{{\underline{\partial}\,}}

\def\Tv{\mbox{T}^{\mbox{v}}}
\def\gR{\hat{g}}
\def\Tstarl{T^{\star_{\ell}}}

\begin{document}

\newcommand{\bm}[1]{\mbox{\boldmath $#1$}}

%Environments:
\newtheorem{definition}{Definition}
\newtheorem{lemma}{Lemma}
\newtheorem{proof}{Proof}
\newtheorem{theorem}{Theorem}
\newtheorem{corollary}{Corollary}
\newtheorem{proposition}{Proposition}

%Definiciones:

%\def\defi{\stackrel{\mbox{\tiny \bf def}}{=}}

\def\botl{\bot_{\ell}}
\def\parallell{\parallel_{\ell}}

\def\Sigmaint{\stackrel{\circ}{\Sigma}}
\def\nn{n{}^{\mbox{\tiny $(2)$}}}
\def\ll{\ell^{\mbox{\tiny $(2)$}}}
\def\ellc{\underaccent{\check}{\bmell}}
\def\bmell{\bm{\ell}}
\def\bmomega{{\bm{\omega}}}
\def\omegac{\underaccent{\check}{\bmomega}}

\def\llprime{\ll{}^{\prime}}
\def\nnprime{\nn{}^{\prime}}

\def\A{{\cal A}}
\def\B{{\cal B}}
\def\C{{\cal C}}
\def\D{{\cal D}}
\def\E{{\cal E}} 
\def\F{{\cal F}}

\def\data{(\gamma_{ab},{\ell}_a,\ll, \Kn_{ab}, \Gamb^{a}_{bc})}
\def\M{{\cal M}}
\def\Ein{{\mbox{Ein}}}
\def\gM{g}
\def\gSigma{\gamma}
\def\nablaM{\nabla}
\def\nablab{{\overline{\nabla}}}
\def\Riemb{\overline{R}}
\def\Gamo{{\stackrel{\circ}{\Gamma}}}
\def\nablao{{\stackrel{\circ}{\nabla}}}
\def\Riemo{{\stackrel{\circ}{R}}}

\def\Kn{K}
\def\etan{{\bm{\eta^{\tiny n}}}}
\def\P{P}
\def\U{U}
\def\Y{\overline{Y}}
\def\nl{ (n \cdot \ell)\, }
\def\Gamb{{\overline{\Gamma}}}

\def\H{{\cal H}}
\def\w{{\omega}}
\def\aa{{a}}
\def\b{{b}}

\def\N{\Sigma}
\def\tr{\mbox{tr} \,}
\def\Sb{S_{b}}
\def\Db{{\mathfrak D}_{b}}
\def\D{{\mathfrak D}}
\def\SN{\Sigma\cap\mathcal{N}}

\def\id{(\Sigma,g,K)}
\def\idfull{(\Sigma,g,K;\rho,{\bf J})}
\def\mots{S}
\def\KVmotsM{\vec{\Pi}}
\def\MCVmotsM{\vec{H}}
\def\MCmotsM{H}
\def\gmots{\gamma}
\def\MCVmotsS{\vec{p}}
\def\MCmotsS{p}
\def\KVmotsS{\vec{\kappa}}
\def\gE{\gamma}
\def\MCVES{\vec{p}_{B}}
\def\MCES{p_{B}}
\def\gl{{g_{0}}}
\def\etal{{\eta_{0}}}
\def\kid{(\Sigma,g,K;N,\vec{Y},\tau)}
\def\R{R}
\def\Rg{{R^{g}}}
\def\Rh{{R^{h}}}
\def\KES{\kappa}
\def\hE{\gamma}
\def\MCES{p}

% A useful Journal macro

\def\Journal#1#2#3#4#5#6{#1, ``#2'', {\em #3} {\bf #4}, #5 (#6).}
\def\JournalPrep#1#2{#1, ``#2'',  In preparation}
\def\Monograph#1#2#3#4{#1, ``#2'', #3 (#4)}
\def\Living#1#2#3#4#5#6#7{#1, #2. Living Rev. Relativity \textbf{#3}, #4 (#5),
URL (cited on #6): 
http://www.livingreviews.org/#7}

% Some useful journal names
\def\JGP{\em J. Geom. Phys.}
\def\JDG{\em J. Diff. Geom.}
\def\CQG{\em Class. Quantum Grav.}
\def\JPA{\em J. Phys. A: Math. Gen.}
\def\PRD{{\em Phys. Rev.} \bm{D}}
\def\GRG{\em Gen. Rel. Grav.}
\def\IJT{\em Int. J. Theor. Phys.}
\def\PR{\em Phys. Rev.}
\def\RMP{\em Rep. Math. Phys.}
\def\MNRAS{\em Mon. Not. Roy. Astr. Soc.}
\def\JMP{\em J. Math. Phys.}
\def\DG{\em Diff. Geom.}
\def\CMP{\em Commun. Math. Phys.}
\def\APP{\em Acta Phys. Polon.}
\def\PRL{\em Phys. Rev. Lett.}
\def\ARAA{\em Ann. Rev. Astron. Astroph.}
\def\ANP{\em Annals Phys.}
\def\AP{\em Ap. J.}
\def\APJL{\em Ap. J. Lett.}
\def\MPL{\em Mod. Phys. Lett.}
\def\PREP{\em Phys. Rep.}
\def\AASF{\em Ann. Acad. Sci. Fennicae}
\def\ZP{\em Z. Phys.}
\def\PNAS{\em Proc. Natl. Acad. Sci. USA}
\def\PLMS{\em Proc. London Math. Soth.}
\def\AIHP{\em Ann. Inst. H. Poincar\'e}
\def\ANYAS{\em Ann. N. Y. Acad. Sci.}
\def\SPJ{\em Sov. Phys. JETP}
\def\PAWBS{\em Preuss. Akad. Wiss. Berlin, Sitzber.}
\def\PPLL{\em Phys. Lett. A }
\def\QJRAS{\em Q. Jl. R. Astr. Soc.}
\def\CR{\em C.R. Acad. Sci. (Paris)}
\def\CP{\em Cahiers de Physique}
\def\NC{\em Nuovo Cimento}
\def\AM{\em Ann. Math.}
\def\APP{\em Acta Physica Polonica}
\def\BAMS{\em Bulletin Amer. Math. Soc}
\def\CPAM{\em Commun. Pure Appl. Math.}
\def\PJM{\em Pacific J. Math.}
\def\ATMP{\em Adv. Theor. Math. Phys.}
\def\PRSA{\em Proc. Roy. Soc. A.}
\def\APPT{\em Ann. Poincar\'e Phys. Theory}
\def\RPM{\em Rep. Math. Phys.}
\def\AHP{\em Annales Henri Poincar\'e}

\def\a{{\alpha}}
\def\h{{h}}

\title{Constraint equations for general hypersurfaces and applications
to shells}

\author{Marc Mars \\
Instituto de F\'{\i}sica Fundamental y Matem\'aticas, IUFFyM,
Universidad de Salamanca,\\
Plaza de la Merced s/n, 37008 Salamanca, Spain \\
marc@usal.es}

\maketitle

\begin{abstract}
Hypersurfaces of arbitrary causal character embedded in a spacetime are studied with the aim
of extracting necessary and sufficient free data on the submanifold suitable for reconstructing the spacetime metric
and its first derivative along the hypersurface. The constraint equations
for hypersurfaces of arbitrary causal character are then computed explicitly in terms
of this hypersurface data, thus providing a framework capable of unifying, and extending,
the standard constraint equations in the spacelike and in the characteristic cases to the general
situation. This may have interesting applications in well-posedness problems more general
than those already treated in the literature. As a simple application of the constraint
equations for general hypersurfaces, we derive the field equations for shells of matter
when no restriction whatsoever on the causal character of the shell is imposed.
\end{abstract}

\section{Introduction}

The Cauchy problem is a fundamental ingredient of General Relativity
(and other geometric theories of gravity) as it allows one to encode
the spacetime information  into the geometry of suitable codimension
one submanifolds and their first order variation. This geometric data 
allows for the  reconstruction of the spacetime (more precisely
the domain of dependence of the data) by solving the field
equations of the theory. This, among many   other reasons, 
makes the study
of spacetime hypersurfaces an important branch of General Relativity.

Solving the evolution problem requires, in one way or another, 
the splitting of the spacetime in terms of a foliation by
hypersurfaces. This can be done either explicitly, as in the ADM or related
formalisms, or implicitly, by constructing appropriate coordinate systems
in which the field equations are solved. The splitting also depends 
on the type of problem under consideration. For instance, in the
standard Cauchy problem the  splitting is performed by a family
of spacelike hypersurfaces and the initial dara consists of the
induced metric $\gamma$ and the second fundamental form $\Kn$.
An important 
property of geometric theories of gravity is that they constitute
a constrained system, in the sense that the initial
data is subject to a set of equations called {\it constraint equations}.
In the spacelike case, they take the standard form
\begin{align}
 2 \rho & \defi 2 G_{\alpha\beta} n^{\alpha} n^{\beta} |_{\N} = R(\gamma) - K_{ab} K^{ab} + K^2, 
\quad \quad K \defi \gamma^{ab} K_{ab}, \label{density0} \\
 - J_{a} & \defi 
G_{\alpha\beta} e^{\alpha}_a n^{\beta} |_{\N} = D_b \left ( K^b_{\,\,a} - K \delta^b_{\,\,a}
\right )  \label{momentum0}
\end{align}
where $D$ is the Levi-Civita covariant derivative of $(\N,\gamma)$ and $R(\gamma)$ its
curvature scalar. $G_{\alpha\beta}$ is the Einstein tensor of the
spacetime, $n^{\alpha}$ is the unit normal used to define $K_{ab}$
and $e^{\alpha}_a$ is the push forward to $\M$ of the coordinate
vector $\partial_a$ in $\N$. This standard splitting of spacetime
into space and time
(i.e. via a folation of spacelike hypersurfaces) is not the only relevant
one. Initial data can also be prescribed on a pair of null hypersurfaces
with a common smooth boundary consisting on a codimension-two spacelike
hypersurface. This is the characteristic initial value problem, and also
gives rise to a well-posed evolution problem \cite{Rendall1990}
in the sense that
suitable initial data (satisfying appropriate constraint
equations) also define a unique spacetime solving the field equations
with this initial data. The constraint equations in the characteristic
case take a very different form than in the spacelike case. 
Most approaches require a 2+1 splitting of the null hypersurface
by spacelike, codimension-one surfaces, and the constraint
equations become a  hierarchical set of ODE along the
degeneration direction of the null hypersurface. 
Several
forms of the constraint equations can be found e.g. in \cite{NullCaseData},
\cite{Rendall1990} or \cite{ChruscielPaetz2012}.

The well-posedness of the characteristic initial value problem has been extended
recently  \cite{nullcone} to the limiting case where the two-surface common 
to the pair of null hypersurfaces degenerates to a point, i.e. when there
is only one null hypersurface which is null everywhere except for 
a conical singularity. The initial data in such a case is similar 
to the characteristic one above, except for the need of a careful
analysis of the conical singularity that arises, see
\cite{nullcone}, \cite{JezierskiChrusciel2012}.
Besides the cases mentioned above, the Einstein field
equations are also well-posed for the so-called
Cauchy-characteristic initial value problem \cite{Winicour}, where
the initial
data is prescribed on a pair of hypersurfaces, one spacelike and
one null with common boundary on a codimension-two surface. 

In view of the very different formulations of the
constraint equations in the spacelike and in the characteristic cases,
a natural question is whether there is any framework capable 
of dealing with both cases at once. More generally, it would be 
interesting to know how do the constraint equations look like for
{\it any} hypersurface in the spacetime. The distinction of hypersurfaces
into spacelike, null or timelike is rather artificial from  a geometric
perspective of submanifolds in a Lorentzian ambient spacetime. The ideal
setting would be a framework where the constraint equation can be written
for an arbitrary hypersurface with no restriction whatsoever in its
causal character (which hence could change along the hypersurface).
Besides its aesthetic appeal, having such  a common framework would
also be of practical interest. First of all, it would
allow us to address the issue of how do the constraint equations 
on spacelike hypersurfaces transform smoothly into the characteristic
constraint equations, which would help 
us clarify the very different nature of the characteristic constraint
equations with respect to the spatial ones. Furthermore, whatever the
final result may be, the set of variables  involved in the general formulation cannot
be the same as in the spacelike case (because those become singular
when the hypersurface becomes null) and so the method would provide
us with alternative expressions both for the standard spacelike 
and for the characteristic constraint equations, and this
may potentially give
new insights into the standard cases as well. Moreover, it is intuitively
clear that the list of well-posed initial value problems discussed
above should not exhaust all the possibilities. From basic causality arguments
one expects that appropriate initial  data prescribed on a hypersurface
everywhere spacelike or null should also give rise to a well-posed
initial value problem (this would correspond, in essence, to a smoothing
of the characteristic or Cauchy-characteristic initial value problem discussed
above). In order to start thinking about such a possibility, it is
necessary to have a framework capable of dealing with the
constraint equations in such a setting, and which defines the types
of variables where the well-posedness problem would be addressed.

It should also be emphasized that physically
relevant hypersurfaces of varying causal
character are much more common than one may think a priori. A list of
interesting examples can be found in the 
Introduction of \cite{MarsSenovilla1993} where the geometry
of arbitrary hypersurfaces in the spacetime was studied. The developments of that paper were focused
in generalizing the matching conditions from the case of constant causal character (well-developed both in the
spacelike and timelike cases \cite{Darmois1927, Lichnerowicz1955, Synge1952} and in the null case \cite{BarrabesIsrael1991})
to the arbitrary case of varying
causal character and this required a better understanding of the geometry of general hypersurfaces in a spacetime. 
Other relevant examples of hypersurfaces
of non-constant causal character 
are the so-called marginally outer trapped tubes (and their
close relatives, the trapping horizons \cite{Hayward1994}, and dynamical/isolated horizons
\cite{AshtekarKrishnan2004}). These are hypersurfaces foliated by codimension-two
spacelike surfaces with one of its null expansions identically vanishing. A priori, these hypersurfaces may have any causal
character. Under appropriate stability and energy conditions, no timelike portion may exist \cite{AMS1,AMS2} but they can still vary
their causal character from spacelike to null, each case having a clear physical interpretation in terms
of the energy flux that crosses the hypersurface. 
Marginally outer trapped tubes are physically very relevant since
they are suitable quasi-local replacements for black hole event horizons, and are analyzed routinely in any 
numerical evolution of ``black hole'' mergers in any collapsing process. 

The aim of this paper is to develop a consistent framework capable of describing the geometry
of arbitrary hypersurfaces in the spacetime in such a way that the constraint equations can be written down in full
generality. The starting point for the construction is based on the results of \cite{MarsSenovilla1993}
(although the presentation will be essentially
self-consistent). It is clear that a fundamental ingredient
of any initial value formulation is the ability of detaching completely the hypersurface from the 
spacetime where it is initially sitting. This is necessary in order to define the data
at the abstract level, without the need of invoking the spacetime for its definition. This will be
the guiding principle of the derivations below. Indeed, the geometry of general spacetime
hypersurfaces  will be studied in detail with the aim of extracting
a set of free geometric data living directly on the submanifold. This
will allow
us to define abstract data and make contact with the spacetime construction via
an appropriate notion of embedding. After this data
is identified, I will derive the constraint equations relating the hypersurface
data with suitable components of the Einstein tensor of the ambient spacetime along
the hypersurface. The Einstein tensor components that can be related to the hypersurface data 
are the normal-tangential component and the normal-transversal components (the precise definitions
will appear below) analogously as in the standard spacelike case. With these identities
at hand, the constraint equations will be promoted to field equations for
matter-hypersurface data at the abstract level, without the need of any embedding
into a spacetime. The constraint equations will be the main
result of this paper and will open up the possibility of studying 
well-posedness issues (particularly in the case of nowhere timelike initial
data sets) in future developments.

As a simple application of the constraint equations for general hypersurfaces
I will obtain the field equations that need to be satisfied for shells
propagating in arbitrary spacetimes, with no restriction on the causal character
of the shell. Recall that a shell arises when two spacetimes with boundary are matched across
their boundaries in such a way that a spacetime with 
continuous metric (in a suitable atlas) is constructed.  When the extrinsic
geometry across the matching hypersurface jumps, this is interpreted as a shell of matter-energy with
support on the hypersurface. It is possible to define an energy-momentum tensor on the shell
which satisfies field equations where the sources are the jumps across the matching
hypersurface of suitable components of the spacetime energy-momentum tensor.
These are the so-called
Israel field equations (also ``shell equations'' or ``surface layer equations'') and
where derived in the case of spacelike or timelike hypersurfaces
first by Lanczos \cite{Lanczos1922, Lanczos1924} and then put in a geometrically clear context by
Israel \cite{Israel1966}. By performing a suitable limit of the equations (in the right variables)
when the spacelike/timelike hypersurface approaches a null limit, these equations were extended
to the case of null hypersurfaces by  Barrab\`es and Israel 
\cite{BarrabesIsrael1991}. The standard method to derive the Israel equations consists in 
using tensor distributions on the spacetime constructed by matching two spacetimes with boundary.
The energy-momentum tensor of the matched spacetime is a distribution which, in general,
has a Dirac delta part supported on the matching hypersurface. This singular part defines the
energy-momentum tensor on the shell and the contracted (distributional) Bianchi identities
lead to the shell equations. This distributional approach can in principle also
be followed in the case of matching hypersurfaces of arbitrary causal character (the distributional setting
for this case was developed in detail in \cite{MarsSenovilla1993}). Nevertheless, having
the constraint equations for arbitrary hypersurfaces at hand, the shell field equations can also
be derived directly by simple subtraction of the constraint equations at each
side of the matching hypersurface.
 Besides its intrinsic simplicity (with no need of using spacetime distributions and
transforming the result back into hypersurface information), this has the advantage that it works even
if the hypersurface data  does not come from any spacetime. This allows us to define shells and 
shell equations  fully independently of the existence of any spacetime where the data is embedded.

The plan of the paper is as follows. In section \ref{Geom}, I will extend the results of
\cite{MarsSenovilla1993} on the geometry of general hypersurfaces. In particular, I will identify the
data that allows one to reconstruct the ambient spacetime metric along the hypersurface (this
leads to the definition of {\bf hypersurface metric data}). Then I will consider the first derivatives
of the spacetime metric along the hypersurface and will extract the corresponding 
free data on the hypersurface (Proposition \ref{solving}). This will lead 
to the definition of {\bf hypersurface data}. The definition of metric hypersurface and hypersurface
data have a built-in gauge freedom tied to the choice of transversal direction used
to define extrinsic
properties of the hypersurface (the so-called rigging vector). The gauge freedom on the hypersurface
data at the abstract level will be studied in detail in Section \ref{gaugeTrans}. In section \ref{constraints},
I will study the
Gauss and Codazzi equations of hypersurfaces in order to write down the normal-tangential and normal-transversal
components of the Einstein tensor in terms of hypersurface data. This will be done first in terms
of a natural connection \cite{MarsSenovilla1993}
on the hypersurface that arises from projection along the rigging
of the spacetime connection onto
the submanifold. This connection,
however geometrically natural in a spacetime setting, has the inconvenience
that it depends on the extrinsic information of the hypersurface data. In a second step, I will rewrite the
constraint equations in terms of a connection that depends solely on the metric hypersurface data and 
in such a way that all the dependence of the extrinsic geometry (the tensor $Y_{ab}$ introduced in the text) 
is fully explicit. This will be the set of equations that I will promote to constraint equations at the 
abstract level. In Section \ref{ThinShells}, I will obtain the shell equations mentioned above by simple
subtraction of the constraint equations obtained in the previous section. During the process,
a symmetric, two-covariant tensor will arise naturally from the equations. This will define
the {\bf energy-momentum tensor} on the shell. In terms of this tensor the shell equations take a very simple 
form. I will conclude with  Section \ref{Conclusions} where a brief summary of results and a discussion of future 
research will be given.

\section{Geometry of general hypersurfaces in an
$(m+1)$-dimensional spacetime.}
\label{Geom}

In this paper manifolds are always connected and paracompact. A 
spacetime is an $(m+1)$-dimensional smooth oriented
manifold $\M$ endowed with 
a symmetric 2-covariant tensor field $\gM$ of Lorentzian signature
$\{-,+, $ $\cdots, + \}$.
The metric $\gM$ will be assumed to be $C^2$.
A ``hypersurface''
is an embedded submanifold of codimension-one, i.e.
a $n$-dimensional smooth manifold $\N$ and an embedding 
$\Phi: \N \rightarrow \M$, where by ``embedding'' we mean
a smooth injective immersion
which is a homomorphism between $\N$ with its manifold topology and
$\Phi(\N)$ with its induced topology as a subset of $\M$. We often
identify $\N$ with $\Phi(\N)$ when necessary.

The first fundamental form of $\N$ is 
$\gamma \defi \Phi^{\star} (\gM)$ of $\N$. 
The signature of $\gamma$ at a given point $p \in \N$ can be either
Euclidean, Lorentzian or of type $\{0,+,+, \cdots, + \}$. In this
paper we deal with arbitrary hypersurfaces and hence we will not assume
that the signature of $\gamma$ remains constant in $\N$.
Since $\gamma$ may be degenerate at some points (or
everywhere), $\N$ inherits, in general, no induced metric  or 
canonical connection from the ambient spacetime. In order to
describe the intrinsic
and extrinsic geometry of $\N$, it is convenient to introduce
an additional structure, namely a spacetime vector field along $\N$ with
is transverse to $\N$ everywhere. Such vector field, called
{\bf rigging} was first
introduced by Schouten \cite{Schouten1954}.

Let $T_{\N} \M$ be
the vector bundle over $\Phi(\N)$ (i.e, the bundle with base $\N$
and fiber at $p \in \N$ the tangent space $T_{\Phi(p)} \M$).
Let $T \, \N$ be the tangent bundle
of $\N$. Identifying $\N$ with $\Phi(\N)$ we can view $T \, \N$
as a vector subbundle of $T_{\N} \M$. The set of smooth sections
on a bundle $(E,\Omega,\pi)$ will be denoted by 
$\Gamma(E)$.

\begin{definition}[Schouten \cite{Schouten1954}]
A {\bf rigging} $\ell$ is a smooth section 
$\ell \in \Gamma (T_{\N} \M)$ satisfying $\ell|_p \not \in T_p \N$ for
all $p \in \N$. 
\end{definition}

An important issue concerning riggings is their existence.
First of all we note that
there always exists a vector bundle $\Tv \, \N$ ($\Tv$ stands
for transverse) over $\N$ such that the vector
bundle decomposition $T_{\N} \M = T \, \N \oplus  \Tv \, \N$
holds. One way of seeing this is by selecting an arbitrary Riemannian
metric $\gR$ on $\M$ (this exists because $\M$ is paracompact, e.g.
\cite{Lee}). At a point $p \in
 \N$ define $\hat{N}_p \N$ as the vector subspace in $T_p \M$
consisting of vectors orthogonal to $T_p \N$ with the metric $\gR$. 
It is immediate that the collection of all $\hat{N}_p \N$, $p \in \N$
defines a vector bundle $\hat{N} \N$ over $\N$ 
and that $T_{\N} \M = T \N  \oplus \hat{N} \N$, which proves the existence
claimed. We note that this construction works in arbitrary codimension.
Note, however, that the existence of the global decomposition
$T_{\N} \M = T \, \N \oplus \Tv \, \N$
fails short of proving the existence of a rigging.
For that it is necessary that a global, nowhere zero section of 
$\Tv \, \N$ exists. The following lemma shows that this happens
if and only if $\N$ is orientable.

\begin{lemma}
Let $\N$ be a hypersurface in $\M$. A rigging $\ell$ exists if and
only if $\N$ is orientable
\end{lemma}

\noindent {\it Proof.} Select one decomposition $T_{\N} \M= T \, \N \oplus
\Tv \,\N$.
If $\N$ is orientable, then there exists a smooth field of
normals $\bm{n}$ (i.e. a smooth field of one-forms on $\N$, nowhere zero and
orthogonal to all tangent vector fields to $\N$). Define $\ell|_p$ as 
the unique vector satisfying $\ell|_p \in \Tv_{p} \N$ and
 $\bm{n}|_p ( \ell|_p ) = 1$. It is obvious that $\ell$ is smooth, nowhere zero
and transverse to $\N$ everywhere, hence a rigging. For the converse,
select the rigging and define, at each point $p \in 
\N$ the one-form $\bm{n}|_p \in \left ( T_p \N \right )^{\bot}
\subset T^{\star}_{p} \M$ (orthogonal in the sense
of dual spaces) satisfying $\bm{n}|_p (\ell|_p ) = 1$ (such $\bm{n}|_p$
exists because $\ell|_p$ is 
not tangent to $\N$ and hence any non-zero one-form
in $\left ( T_p \N \right )^{\bot}$ when applied to $\ell|_p$ gives a non-zero value). It is immediate to check that ${\bm n} : \N \rightarrow
T^{\star} \N$, defined as ${\bm n} (p) \defi \bm{n}|_p$ gives a smooth
field of normals, and hence $\N$ is orientable. \ep

\vs

In view of this lemma, we will from now one assume that all hypersurfaces
are orientable unless contrarily specified.

\vs

It is clear that the choice of rigging is highly non-unique, and we will have to
deal with this freedom later. Nevertheless, a rigging $\ell$ allows
for a decomposition \cite{MarsSenovilla1993} $T_p \M = \langle \ell|_p \rangle
\oplus T_p \N$, where $\langle \ell|_p \rangle$ is the vector
subspace of $T_p \M$ generated by $\ell|_p$. 
Given $V \in T_p M$ we define a scalar $V^{\botl}$ and
a vector $V^{\parallell} \in T_p \N$ via the decomposition
$V = V^{\botl} \ell|_p + V^{\parallell}$.  Given a
section $V \in \Gamma (T_{\N} \M)$ this decomposition
defines a scalar $V^{\botl} : \N \longrightarrow \mathbb{R}$
and $V^{\parallell} \in \Gamma( T \, \N)$. These definitions obviously
depend on the choice of rigging $\ell$. 

The decomposition $T_p \M = \langle \ell|_p \rangle \oplus T_p \N$
induces a decomposition of the dual space $T^{\star}_p \M = \Tstarl_p \, \N 
\oplus N_p \N$, where $\Tstarl_p \, \N =
\langle \ell|_p \rangle^{\bot} \subset T^\star_p \M$
and $N_p \, \N = ( T_p \N )^{\bot}$. The latter is the normal space
to $\N$ at $p$, and its elements are normal one-forms to $\N$.
Note that $N_p \N$ is independent of the rigging, while $\Tstarl_p
\, \N$ is not. It is also clear that the collection of $\Tstarl_p \,
\N$, $p \in \N$, defines a vector bundle over
$\N$ , denoted by $\Tstarl \N$. The same
occurs for the collection $N_p \, \N$, which 
defines the normal bundle  $N \N$. This bundle
is independent of the choice of rigging. However,
given $\ell$, we define $\bm{n}$
as the unique normal one-form $\bm{n} \in \Gamma (N \N)$
satisfying $\bm{n} (\ell) = 1$. Despite the fact that
$\bm{n}$ depends on $\ell$, we will not make this dependence explicit
in the symbol in order not to make the notation cumbersome. We will
do the same for several other objects defined below. 
We also note that, in the same way as we have identified
$\N$ with its image we have also identified $T \N$, the tangent
bundle of $\N$ as an abstract manifold, with the vector subbundle
$T \N \subset T_{\N} \M$. The precise meaning of an object in such a space
will be either clear from the context, or made explicit.

In this paper we will often use index notation. To that aim,
let  $\{ \hat{e}_a \}$ $a=1,\cdots, m$ be a basis of
of $T \N$.  By definition, this means a set of $m$ smooth
sections $\hat{e}_a \in \Gamma ( T \, \N)$ such that for all $p \in \N$
$\{ \hat{e}_a |_p \}$ is a basis of $T_p \N$ \footnote{In general, no such global
basis exists, and we would need to work with bases defined on each element
of a suitable open
cover of $\N$. Since all the expressions below will be tensorial (unless
explicitly stated), there
is no loss of generality in working as if the global basis did exist. This
difficulty is general to the use of index notation and it is both
well-understood and harmless. An alternative is to view
indices in the sense of the abstract index notation of Penrose}.  The set of $m+1$ vectors
$\{ e_a , \ell \}$, where $e_a = \Phi_{\star}(e_a)$ is clearly
a basis of $T_{\N} \M$. Its dual basis is composed by 
$\{ \bmomega^a, \bm{n} \}$, where the $n$ one-forms $\{ \bmomega^a \}$ 
are defined by $\bmomega^a  ( e_ b ) = \delta^a_b$, $\bmomega^a  (\ell)
= 0$. It is clear that 
$\bmomega^a $ also depends on the choice of rigging. By construction
$\{\bm{\omega^a} \}$ is a basis of
$\Tstarl_p \N$. It is also immediate that
the pull-back of $\omega^a $ to $T^{\star} \N$, i.e.
$\omegac^a \defi \Phi^{\star} (\bmomega^a)$,
defines a basis of this space.

Everything we have said so far is independent of the existence of 
a metric $\gM$ in the ambient manifold $\M$. Assume now that $\M$
is endowed with a metric of Lorentzian signature $\gM$. We can then 
define the scalar $\ll = \gM (\ell, \ell)$ and the
one-form $\bmell \defi \gM(\ell, \cdot )$. Pulling this back
to $\N$, we obtain a one-form  $\ellc \defi \Phi^{\star} (\bmell)$
which can be decomposed in the basis $\{ \omegac^a \}$
as  $ \ellc = \ell_a \omegac^a$, for certain coefficients
$\ell_a$. An alternative (and equivalent) definition of $\ell_a$
is $\ell_a = \gM(\ell,e_a)$. 
If, as before, we denote by $\gamma$, the pull-back on $\N$ of the ambient
metric  $\gM$, this tensor may be degenerate (at certain points, or nowhere,
or everywhere). However, it must be the case that the square $(m+1)$-matrix
\begin{equation*}
\begin{bmatrix}
\gamma_{ab} & \ell_a \\
\ell_b & \ll 
\end{bmatrix}
\end{equation*}
has Lorentzian signature at every point $p \in \N$ (because
this is simply the matrix representation of the ambient
metric $\gM$ in the basis $\{ e_a, \ell\}$). This suggests the following
definition, where everything refers to $\N$ as an abstract manifold, not
embedded in any ambient spacetime.

\begin{definition}
\label{hypmetdata}
A smooth $m$-dimensional manifold $\N$, a symmetric tensor $\gamma_{ab}$,
a one-form $\ell_a$ and a scalar $\ll$ define a 
{\bf hypersurface metric data} set provided the
square $(m+1)$-matrix 
\begin{equation*}
\mathbb{A} \defi 
\begin{bmatrix}
\gamma_{ab} & \ell_a \\
\ell_b & \ll 
\end{bmatrix}
\label{gr}
\end{equation*}
has Lorentzian signature at every point $p \in \N$.
\end{definition}

\noindent {\bf Remark \arabic{remark}. \addtocounter{remark}{1}}
%Note that $\ll$ in this definition is a scalar which exists on its own,
%not as the square of any quantity (since there is no embedding, there
%is no rigging vector either). In particular $\ll$ is not assumed to have
%any sign a priori. 
Note that, by definition, $\ell_a$, $\ll$ cannot
vanish simultaneously 
at any point in any hypersurface metric data.

\vs

\noindent {\bf  Remark \arabic{remark}. \addtocounter{remark}{1}}
The most interesting case for gravity arises
when hypersurfaces are
embedded in a spacetime, i.e. in  a manifold with a metric of Lorentzian singature. Nevertheless, 
the signature of the ambient manifold will be used essentially
nowhere below. In fact, all the developments of this paper
can be generalized with very
minor changes to hypersurfaces embedded in ambient manifolds endowed with
a metric of arbitrary (non-degenerate) signature.

\vs

Given hypersurface metric data, we can define immediately
a symmetric two-contravariant tensor $\P^{ab}$, a
vector $n^a$ and a scalar $\nn$ in $\N$, as the unique tensors
satisfying the tensor equations on $\N$,
\begin{eqnarray}
\P^{ab} \gamma_{bc} + n^a \ell_b = \delta^a_b, \label{EqP1} \\
\P^{ab} \ell_b + \ll n^a = 0, \label{EqP2} \\
n^a \ell_a  + \nn \ll = 1,   \label{EqP3} \\ 
\gamma_{ab} n^b  + \nn \ell_a = 0.\label{EqP4}
\end{eqnarray}
Existence and uniqueness of $\P^{ab}$, $n^a$ and $\nn$ is immediate by noticing
that these equations  can be put in matrix form as
\begin{equation}
\begin{bmatrix}
P^{ab} & n^a \\
n^b & \nn 
\end{bmatrix} \defi
\begin{bmatrix}
\gamma_{ab} & \ell_a \\
\ell_b & \ll 
\end{bmatrix}^{-1}.
\label{inverse}
\end{equation}
Given hypersurface metric data, we will always
define $P^{ab}$, $n^a$ and $\nn$ as the solutions of (\ref{EqP1})-(\ref{EqP4})
unless contrarily indicated. Furthermore, we define the
vector field $\hat{n} \defi n^a \hat{e}_a \in \Gamma(T \Sigma)$.

We want to think of hypersurface metric data $\{ \N,\gamma_{ab},\ell_a,\ll\}$ 
as an abstract collection of objects, defined independently
of any spacetime and any embedding. To make contact with the previous
discussion, the following definition is required.
\begin{definition}
A hypersurface metric data $\{\N,\gamma,\ellc,\ll\}$ is
{\bf embedded} in a spacetime $(\M,\gM)$ if there
exists an embedding $\Phi : \N \rightarrow \M$ and a rigging
vector $\ell$ such that, with $\bmell = g (\ell,\cdot)$,
\begin{eqnarray*}
\gamma = \Phi^{\star} (\gM ), \quad \quad \ellc = \Phi^{\star} (\bmell),
\quad \quad \ll = \Phi^{\star} (\gM (\ell,\ell)).
\end{eqnarray*}
\end{definition}
The following lemma, gives the relationship between $n^a, \nn$ and $P^{ab}$
with the ambient geometry when the hypersurface metric data
is embedded.
\begin{lemma}
Let $\{\N,\gamma,\ellc,\ll\}$ be
embedded hypersurface metric data with embedding
$\Phi$, spacetime $(\M,\gM)$ and rigging vector $\ell$. 
Let $\{ \hat{e}_a \}$ be a basis of $T \N$ and $e_a = \Phi_{\star} (\hat{e}_a)$. 
Then, the vectors
$n \defi g^{-1} (\bm{n}, \cdot)$ ($g^{-1}$ is the inverse tensor of $g$)
$\omega^a \defi 
g^{-1} (\bmomega^a, \cdot)$ and the one-forms $\bm{e}_a \defi
g (e_a, \cdot)$, $\bmell = g (\ell,\cdot)$ can be decomposed as
\begin{eqnarray}
n = n^a e_a + \nn \ell, \label{n}\\
\bmell = \ell_a \bmomega^a + \ll \bm{n}, \label{bmell} \\
\bm{e}_a = \gamma_{ab} \bmomega^b + \ell_a \bm{n}, \label{bme} \\
\omega^a = \P^{ab} e_b + n^a \ell. \label{omegaa}
\end{eqnarray}
 \end{lemma}
\noindent {\it Proof.} For (\ref{n}), we only need to check that
$n$ defined by this formula is normal to $\N$ and that it satisfies
$\gM(n,\ell)=1$.
\begin{align*}
\gM(n,e_a) & = n^b  \gamma_{ab} + \nn \ell_a = 0, \quad \quad \quad \quad & \mbox{by }
(\ref{EqP4}),
\\
 \gM(n,\ell) & = n^a \ell_a + \nn \ll = 1, \quad \quad \quad \quad & \mbox{by }
(\ref{EqP3}),
\end{align*}
where we used $\gM( e_a, e_b ) = \gamma_{ab}$ and
$g(e_a, \ell) = \ellc(\hat{e}_a ) = \ell_a$.
Expression (\ref{bmell}) follows from the immediate facts that
$(\ell_a \bmomega^a + \ll \bm{n} ) (e_b) = \ell_a$ and
$(\ell_a \bmomega^a + \ll \bm{n} ) (\ell) = \ll$.
For (\ref{bme}), we should check whether 
$\bm{e}_a (e_b) =  \gM( e_a, e_b ) = \gamma_{ab}$
and $\bm{e}_a (\ell) = \bmell(e_a) = \ell_a$. Both are 
immediate. For (\ref{omegaa}) we need to check
$\gM (\omega^a,e_b) = \delta^a_{b}$ and $\gM(\omega^a,\ell) = 0$. Indeed
\begin{align*}
 \gM(\omega^a,e_b) & = P^{ac} \gamma_{cb} + n^a \ell_b = \delta^a_b, \quad \quad
\quad \quad  & \mbox{by }
(\ref{EqP1}), \\
 \gM(\omega^a,\ell) & = P^{ab} \ell_b  + n^a \ll = 0, \quad \quad \quad \quad& \mbox{by }
(\ref{EqP2}). 
\end{align*}
\ep

\vs

A consequence of this lemma is that, for embedded 
hypersurface metric data, we have $\hat{n} = n^{\parallell}$.
It also implies that, for embedded hypersurface metric data,
the quantities $P^{ab}$, $n^a$ and $\nn$ can also be calculated from the
expressions
\begin{eqnarray}
P^{ab} = \gM^{-1}(\bmomega^a, \bmomega^b), \quad n^a = \gM^{-1}(\bm{n},\bmomega^a), 
\quad \nn = \gM^{-1} (\bm{n},\bm{n}). \label{ginv}
\end{eqnarray}
In this context, these expressions
could have been obtained also from the fact that the matrix 
components of $g^{-1}$ in the basis $\{\bmomega^a, \bm{n} \}$
is precisely 
\begin{equation*}
\begin{bmatrix}
P^{ab} & n^a \\
n^b & \nn 
\end{bmatrix}.
\end{equation*}

The following simple lemma allows us to reconstruct a vector
$V^a$ from the one-form $V_a \defi \gamma_{ab} V^b$ and the contraction
$V^b \ell_b$, and will be used many times below.
\begin{lemma}
\label{raising}
Let $Z_a$ and $W$ be given. There exists a vector $V^a$
satisfying $V^a \ell_a = W$ and $\gamma_{ab} V^b = Z_a$ if and only if
\begin{eqnarray}
n^b Z_b + \nn W = 0.
\label{cond}
\end{eqnarray}
Moreover, the solution is unique and reads
\begin{eqnarray}
V^a = P^{ab} Z_b + n^a W. \label{sol}
\end{eqnarray}
\end{lemma}
\noindent {\it Proof.}  Assume (\ref{cond}). Let us check that (\ref{sol})
solves the two equations $V^a \ell_a = W$, $\gamma_{ab} V^b = Z_a$. Indeed,
using (\ref{EqP1})-(\ref{EqP4}) one has
\begin{eqnarray}
\gamma_{ab} V^b = \gamma_{ab} P^{bc} Z_c + 
n^b \gamma_{ab} W =
\left ( \delta^c_a - n^c \ell_a \right ) Z_c - \nn \ell_a W 
= Z_a - \ell_a
\left ( n^c Z_c + \nn W \right ) =  
Z_a, \label{first} \\
V^a \ell_a = P^{ab} Z_b \ell_a +  n^a \ell_a W =
- \ll n^b Z_b +
\left ( 1 - \nn \ll \right ) W = 
 W -  \ll \left ( n^b Z_b + \nn W \right ) = W.
\label{second}
\end{eqnarray}
To show necessity, let $V^a$ solve the equations
$V^a \ell_a = W$ and $V^b \gamma_{ab} = Z_a$. Multiplying the second by $P^{ac}$
yields
\begin{eqnarray*}
Z_a P^{ac} = V^b \gamma_{ab} P^{ac} = V^b \left ( \delta^c_b - 
n^c \ell_b \right ) = V^c -  n^c W
\end{eqnarray*}
Thus (\ref{sol}) is the only possible solution (by the way, this proves
the claim of uniqueness). Now, the calculations (\ref{first}) 
and (\ref{second}) are  still valid. The last equality in
both expressions implies (\ref{cond})
because it cannot happen that $\ell_a =0$ and $\ll=0$ simultaneously.
\ep

\vs 
Let us now denote the  Levi-Civita covariant derivative of $(\M,\gM)$ by
$\nabla$. Given two vectors $X,Y \in \Gamma (T \N)$ we define 
\begin{eqnarray}
\nablab_X Y & \defi & \left (\nabla_X Y  \right)^{\parallell}, 
\label{GambAmb}\\
\Kn \left (X,Y\right ) & \defi & -  \left ( \nabla_X Y
\right )^{\botl}. \label{KnAmb}
\end{eqnarray}
It is immediate to check \cite{MarsSenovilla1993}
that $\nablab$ defines a torsion-free
covariant derivative on $\N$ and that $\Kn (X,Y)$ is 
a symmetric tensor. $\Kn$ only depends on $\bm{n}$, as the following
alternative expression implies,
\begin{eqnarray*}
\Kn (X,Y) = - \gM (\bm{n}, \nabla_X Y ) = 
\left ( \nabla_X \bm{n} \right ) ( Y),
\end{eqnarray*}
where $\bm{n}$ in the last expression is any smooth  extension of $\bm{n}$
off $\N$. This expression shows that 
$\Kn$ is the second fundamental form of $\N$
with respect to the normal $\bm{n}$. In the Riemannian case,
this tensor captures the extrinsic geometry of the submanifold. However,
for general hypersurfaces, the normal vector $n$ is tangent
to $\N$ at points where the hypersurface is degenerate. Thus, $\Kn$
gives no extrinsic information on the geometry of $\N$ at those points. In 
the following, we will identify a suitable tensor that will 
encode the information on the extrinsic geometry of $\N$. To that aim,
let us introduce the Christoffel symbols $\Gamb^{a}_{bc}$ 
of the connection $\nablab$ in a given basis $\{ \hat{e}_a \}$. 
Following \cite{MarsSenovilla1993} we also define 
a one-form $\varphi_a$ and an endomorphism  $\Psi^{a}_{\,\,\,b}$ by 
\begin{align}
  \varphi_a & = - 
\left ( \nabla_{e_a}  \bm{n} \right ) ( \ell ), \label{varphiAmb}\\
 \Psi^{a}_{\,\,\,b} & = \bmomega^b \left ( \nabla_{e_a} \ell \right ).
\label{PsiAmb}
\end{align}
Note that $\Psi^a_{\,\,\,b}$ are simply the coefficients of
$\left ( \nabla_{e_a} \ell \right )^{\parallell}$ in the basis
$\{ e_{b} \}$. The definitions above imply \cite{MarsSenovilla1993}
\begin{eqnarray}
\nabla_{e_a} e_b  & = &  - \Kn_{ab}  \ell + \Gamb^{c}_{ba}
e_c, \label{covder1} \\
\nabla_{e_a} \ell & = &  \varphi_a  \ell + \Psi^b_{\,\,\,a} e_b, \label{covder2} \\
\end{eqnarray}
These equations are equivalent to the following, written in the
dual basis $\{ \bmomega^a,\bm{n}\}$  of $\{e_a,\ell\}$m 
\begin{eqnarray}
\nabla_{e_a} \bm{n} & = &  - \varphi_a  \bm{n}  + \Kn_{ab}  \bmomega^b, 
\label{covder3}  \\
\nabla_{e_a} \bmomega^b & = &  - \Psi^b_{\,\,\,a} \bm{n} 
- \Gamb^b_{ca} \bmomega^c. \label{covder4}
\end{eqnarray}
As before, we want to identify the minimal set of quantities on
$\N$ that allows us to define hypersurface data in a detached form
from the spacetime and the embedding. To that aim, we will first
obtain which compatibility conditions must satisfy the fields
$\Kn_{ab}, \Gamb^c_{ab}, \varphi_a, \Psi^b_{\,\,\,a}$ when defined via
(\ref{covder1})-(\ref{covder2}).
By finding the general solution of those compatibility
condition we will be able to identify the free data on the hypersurface that
will encode the extrinsic information of the embedding.

The compatibility conditions arise from the fact that the connection
on the ambient manifold is metric and torsion-free. Denoting by $\partiald_a$ 
the directional derivative along $e_a$, we have
\begin{eqnarray*}
& & \partiald_a \gamma_{bc} = \nabla_{e_a} \gM(e_b,e_c) =
\bm{e}_b \left ( \nabla_{e_a} e_c \right ) + 
\bm{e}_c \left ( \nabla_{e_a} e_b \right ) = 
- \Kn_{ac} \ell_b + \Gamb^{d}_{ac} \gamma_{ab} 
- \Kn_{ab} \ell_c + \Gamb^{d}_{ab} \gamma_{cd} , \\
& & \partiald_a \ell_b  = \nabla_{e_a} \gM(e_b,\ell) =
\bm{e}_b (\nabla_{e_a} \ell ) + 
\bmell (\nabla_{e_a} e_b ) = \varphi_a \ell_b + \Psi^{c}_{\,\,\,a} \gamma_{bc}
- \Kn_{ab} \ll + \Gamma^c_{ab} \ell_c, \\
& & \partiald_a \ll = 2 \bmell ( \nabla_{e_a} \ell ) = 2 \varphi_a \ll
+ 2 \Psi^{b}_{\,\,\,a} \ell_{b}.
\end{eqnarray*}
Thus, the compatibility equations take the following tensorial form
\begin{eqnarray}
\nablab_a \gamma_{bc} +  \ell_b \Kn_{ac} + \ell_c \Kn_{ab} 
& = &  0,\label{comp1} \\
\nablab_a \ell_b - \varphi_a \ell_b
+ \ll \Kn_{ab} - \gamma_{bc} \Psi^c_{\,\,\,a} & = &  0, 
\label{comp2} \\
- \frac{1}{2} \nablab_a \ll + 
\Psi^b_{\,\,\,a} \ell_b 
+ \ll \varphi_a & = &  0.
\label{comp3} 
%\nablab_a P^{bc} + 
%\left ( n^b \Psi^c_{\,\,\,a} + n^c \Psi^b_{\,\,\,a} \right )  & = &  0, \label{comp4} \\
%\nablab_a n^b + \varphi_a n^b- P^{bc} \Kn_{ac} + 
%\nn \Psi^b_{\,\,\,a}  & = &  0, \label{comp5} \\
%K_{ab} n^b - \nn \varphi_a - \frac{1}{2}
%\nablab_a \nn & = &  0 \label{comp6}
\end{eqnarray}
Given hypersurface metric data, we can consider these equations
as equations for the unknowns $\{\Gamb^c_{ab}$, $\Kn_{ab}$, $\Psi^{b}_{\,\,\,a}$,
$\varphi_a \}$. The following Proposition provides the general 
solution in terms of a free symmetric two-covariant tensor on $\N$.

\begin{proposition}
\label{solving}
Let $\{\N,\gamma,\ellc,\ll\}$ be hypersurface metric data and
$Y_{ab}$ an arbitrary symmetric tensor on $\N$. 
Let $\bm{F}$ be the two-form $\bm{F} \defi \frac{1}{2} d \ellc$
%$F_{ab} \defi  \frac{1}{2} \left ( \partial_a \ell_b - \partial_b \ell_a \right )$}
and define, in any coordinate basis,
\begin{eqnarray}
\Gamb^{c}_{ab} & \defi &
\frac{1}{2} P^{cd} \left ( \partial_a \gamma_{bd} 
+ \partial_b \gamma_{ad} - \partial_d \gamma_{ab} \right )
+ n^c  \left ( - Y_{ab} + \frac{1}{2} \left (\partial_a \ell_b
+ \partial_b \ell_a \right )  \right ), \label{GambProp}\\
\Kn_{ab} & \defi & 
\nn Y_{ab} + \frac{1}{2} {\pounds}_{\hat{n}} \gamma_{ab} + \frac{1}{2}
\left ( \frac{}{} \ell_a \partial_b \nn + \ell_b \partial_a \nn \right ), 
\label{KnProp} \\
\varphi_{a} &\defi& \frac{1}{2} \nn \partial_a \ll + n^b \left (Y_{ab}
+ F_{ab} \right ) , \label{varphiProp} \\
\Psi^{b}_{\,\,\,a} &\defi&  P^{bc} \left ( Y_{ac} + F_{ac}
\right ) + \frac{1}{2} n^b  \partial_a \ll, \label{PsiProp} 
\end{eqnarray}
where $\pounds$ denotes the Lie derivative.
Then $\Gamb^b_{ab}$ defines a torsion-free connection on $\N$
and $\{ \Gamb^{b}_{ab}, \Kn_{ab}, \Psi^{b}_{\,\,\,a},\varphi_a \}$ 
solves the compatibility equations (\ref{comp1})-(\ref{comp3}).
Conversely, any solution of these equations can be written
in this form for some symmetric tensor $Y_{ab}$. In either case,
the tensor $Y_{ab}$ satisfies the identity
\begin{eqnarray}
Y_{ab} = \frac{1}{2}  \left ( \nablab_a \ell_b + \nablab_b \ell_a \right )
+ \ll \Kn_{ab}. \label{YProp}
\end{eqnarray}
\end{proposition}

\noindent {\it Proof.} We can apply Lemma \ref{raising} to equations (\ref{comp2})
and (\ref{comp3}) with $V^b \rightarrow \Psi^b_a$,
$Z_b \rightarrow  \nablab_a \ell_b - 
\varphi_a \ell_b 
+ \ll \Kn_{ab}$ and $W \rightarrow
\frac{1}{2} \nablab_a \ll
- \ll \varphi_a$. This gives, on the one hand, an explicit
expression for $\Psi^{b}_{\,\,\,a}$, namely
\begin{eqnarray}
 \Psi^{b}_{\,\,\,a} = P^{bc} \nablab_a \ell_c
+ \ll
P^{bc} \Kn_{ca} + \frac{1}{2} n^b \nablab_a \ll, \label{Psi} 
\end{eqnarray}
and, on the other, the compatibility equation (\ref{cond}) which reads
\begin{eqnarray*}
n^b \left ( 
  \nablab_a \ell_b - \varphi_a \ell_b 
+ \ll \Kn_{ab} 
\right ) + \nn 
\left (  \frac{1}{2} \nablab_a \ll
- \ll \varphi_a \right ) = 0.
\end{eqnarray*}
Recalling 
(\ref{EqP3}), this equation gives an explicit expression for
$\varphi_a$, namely
\begin{eqnarray}
\varphi_a = n^b \left ( 
  \nablab_a \ell_b 
+ \ll \Kn_{ab} 
\right ) + \frac{1}{2} \nn \nablab_a \ll. \label{varphi} 
\end{eqnarray}
Equations (\ref{comp2}) and (\ref{comp3}) are therefore equivalent to
(\ref{Psi}) and (\ref{varphi}). Thus, we only need to solve (\ref{comp1})
in terms of free data.  Assume first that we are given a collection
$\{\gamma_{ab}, \ell_a, \ll, \Kn_{ab},
\Gamb^{a}_{bc} \}$ satisfying (\ref{comp1}).
Let us define
a symmetric tensor $Y_{ab}$ by
\begin{eqnarray}
Y_{ab} \defi \frac{1}{2}  \left ( \nablab_a \ell_b + \nablab_b \ell_a \right )
+ \ll \Kn_{ab}. \label{defY}
\end{eqnarray}
We want to determine $\Kn_{ab}$ and $\Gamb^{c}_{ab}$ 
in terms
of $\{\gamma_{ab},\ell_a,\ll,Y_{ab}\}$. 
Expanding the covariant derivative in (\ref{comp1}), we get (working
in a coordinate basis)
\begin{eqnarray*}
  \partial_a \gamma_{bc}  - \Gamb^{d}_{ba} \gamma_{dc} - \Gamb^{d}_{ca} \gamma_{bd} = - \ell_b \Kn_{ac} - 
\ell_c \Kn_{ab}.
\end{eqnarray*}
Now, take the three cyclic permutations of these
equations and subtract the third one to the sum of the other two. The
result is
\begin{eqnarray}
\Gamb^{d}_{ab} \gamma_{dc} = 
\frac{1}{2} \left (
\partial_a \gamma_{bc} + \partial_{b} \gamma_{ca} - 
\partial_{c} \gamma_{ab} \right ) + \ell_c \Kn_{ab}
\defi Z_{cab} \label{ExpGambgam}
\end{eqnarray}
The definition of $Y_{ab}$ (\ref{defY}) implies 
\begin{eqnarray}
\Gamb^{c}_{ab} \ell_c =
\frac{1}{2} \left ( \partial_a \ell_b + \partial_b \ell_a \right ) -
Y_{ab} + \ll \Kn_{ab} \defi W_{ab} \label{ExpGambell}
\end{eqnarray}
We therefore have expressions for the contraction of $\Gamb^{c}_{ab}$
with the first fundamental form and with $\ell_c$. We can apply
Lemma \ref{raising} with $Z_c \rightarrow Z_{cab}$ and $W \rightarrow W_{ab}$.
This gives an explicit solution for 
$\Gamb^{c}_{ab}$
and a compatibility condition. The expression for $\Gamb^{c}_{ab}$, namely
$\Gamb^{c}_{ab} = P^{cd} Z_{dab} + n^c W_{ab}$, gives
\begin{eqnarray}
\Gamb^{c}_{ab} & = &  
\frac{1}{2} P^{cd} \left ( \partial_a \gamma_{bd} 
+ \partial_b \gamma_{ad} - \partial_d \gamma_{ab} \right )
+ n^c  \left ( - Y_{ab} + \frac{1}{2} \left (\partial_a \ell_b
+ \partial_b \ell_a \right )  \right )
+ \left ( P^{cd} \ell_{d} + n^c \ll \right )  \Kn_{ab}  \nonumber \\
& = & 
\frac{1}{2} P^{cd} \left ( \partial_a \gamma_{bd} 
+ \partial_b \gamma_{ad} - \partial_d \gamma_{ab} \right )
+ n^c  \left ( - Y_{ab} + \frac{1}{2} \left (\partial_a \ell_b
+ \partial_b \ell_a \right )  \right ),
\label{defGam}
\end{eqnarray}
where in the last equality we used (\ref{EqP2}).
The compatibility condition $n^c Z_{cab} + \nn W_{ab} =0$ reads
explicitly
\begin{eqnarray}
0 & = & 
\frac{1}{2} n^c \left ( \partial_a \gamma_{bc} + 
\partial_b \gamma_{ac} - \partial_c \gamma_{ab} \right ) 
+ n^c \ell_c \Kn_{ab} + \frac{1}{2} \nn \left ( \partial_a \ell_b 
+ \partial_b \ell_a \right ) + \nn \ll \Kn_{ab} - \nn Y_{ab} \nonumber \\
& = &
\frac{1}{2} n^c \left ( \partial_a \gamma_{bc} + 
\partial_b \gamma_{ac} - \partial_c \gamma_{ab} \right ) 
+ \frac{1}{2} \nn \left ( \partial_a \ell_b 
+ \partial_b \ell_a \right ) + \Kn_{ab} - \nn Y_{ab},
 \label{compatbis}
\end{eqnarray}
where in the second equality we used (\ref{EqP3}). To elaborate this
expression we note the identity
\begin{align*}
n^c \partial_a \gamma_{bc} + \nn \partial_a \ell_b & =
\partial_a \left ( n^c \gamma_{bc} \right ) - \gamma_{bc} \partial_a n^c
+ \nn \partial_a \ell_b = 
\partial_a \left ( - \nn \ell_b \right ) + \nn \partial_a \ell_b 
- \gamma_{bc} \partial_a n^c =\\ 
& = 
- \ell_b \partial_a \nn - \gamma_{bc} \partial_a n^c,
\end{align*}
where (\ref{EqP4}) has been used. Inserting this
into (\ref{compatbis}) and recalling the 
expression in coordinates for the Lie derivative 
$\pounds_{\hat{n}} \gamma_{ab} = n^c \partial_c \gamma_{ab}
+ \gamma_{ac} \partial_b n^c + \gamma_{bc} \partial_a n^c$, we find
\begin{eqnarray}
\Kn_{ab} = \nn Y_{ab} + \frac{1}{2} \pounds_{\hat{n}} \gamma_{ab} + \frac{1}{2}
\left ( \frac{}{} \ell_a \partial_b \nn + \ell_b \partial_a \nn \right ). 
\label{defKn}
\end{eqnarray}
Expressions (\ref{defGam}) and (\ref{defKn}) give
$\Gamb^c_{ab}$ and $K_{ab}$ explicitly in terms of $Y_{ab}$. Moreover,
$Y_{ab}$ is free data because if we {\it define} 
$\Gamb^{c}_{ab}$ and
$K_{ab}$ in terms of an arbitrary symmetric tensor $Y_{ab}$
through expressions (\ref{defGam})-(\ref{defKn}),
then it is immediate to check that $\Gamb^{c}_{ab}$ is a torsion-free
connection and, in addition,
both equation (\ref{comp1}) and expression (\ref{defY}) (which is now
an equation) are identically satisfied with the covariant derivative 
$\nablab$ defined in terms of the connection $\Gamb^{c}_{ab}$.
Indeed, Lemma \ref{raising} implies that expressions (\ref{defGam})-(\ref{defKn})
are equivalent to (\ref{ExpGambgam})-(\ref{ExpGambell}) and, from the
latter, equations (\ref{comp1}) and (\ref{defY}) follow at once.

To complete the proof, we only need to notice that, irrespective
of whether $Y_{ab}$ is defined by (\ref{defY}) or $\Gamb^{c}_{ab}$
is defined as in (\ref{defGam}) the following holds
\begin{eqnarray}
\nablab_a \ell_b = 
\frac{1}{2} \left ( \nablab_a \ell_b +  \nablab_b \ell_a \right ) + 
\frac{1}{2} \left ( \nablab_a \ell_b -  \nablab_b \ell_a \right ) =
Y_{ab} - \ll \Kn_{ab} + F_{ab} \label{nablabell}
\end{eqnarray}
and (\ref{Psi}), (\ref{varphi}) become, respectively,
(\ref{PsiProp}) and (\ref{varphiProp}). \ep

\vs The following corollary will be useful later

\begin{corollary}
With the same hypothesis and nomenclature as in Proposition
\ref{solving},
\begin{eqnarray*}
n^b K_{ab} = \nn n^b \left ( Y_{ab} + F_{ab} \right ) + \frac{1}{2} 
\left ( \partial_a \nn + ({\nn})^2 \partial_a \ll \right ).
\end{eqnarray*}
\end{corollary}

\vs 

\noindent {\it Proof.}
We only need to contract (\ref{KnProp}) with $n^b$, which in 
particular involves $n^b {\pounds}_{\hat{n}} \gamma_{ab}$. Since 
${\pounds}_{\hat{n}} \hat{n} =0$, we have
\begin{align*}
n^b {\pounds}_{\hat{n}} \gamma_{ab}  = & {\pounds}_{\hat{n}} \left ( n^b \gamma_{ab}
\right )  = - {\pounds}_{\hat{n}} \left (\nn \ell_a \right ) =
- ( n^b \partial_b \nn) \ell_a - \nn 
{\pounds}_{\hat{n}} \left (\ell_a \right )  = 
- (n^b \partial_b \nn ) \ell_a + \nonumber \\
& + \nn \left ( - 2 n^b F_{ba} - \nablab_a
\left ( n^b \ell_b \right ) \right ),
\end{align*}
where in the third equality we used the property
${\cal  L}_{Y} \bm{\alpha} = i_{Y} d \bm{\alpha} + 
d ( i_{Y} \bm{\alpha})$ valid for any differential form $\bm{\alpha}$.
Using this expression in the contraction of $n^b$
with (\ref{KnProp}) and
recalling $n^b \ell_b = 1 - \nn\ll$,  the corollary follows directly.
\ep 

\vs

Following the idea of defining data on $\N$ as detached from any
ambient geometry, we put forward the following definition:
\begin{definition}[Hypersurface data]
A five-tuple
$\{\N,\gamma_{ab},\ell_{a},\ll,Y_{ab}\}$ where
$\{\N,\gamma_{ab},\ell_{a},\ll \}$ is hypersurface metric data and
$Y_{ab}$  is a symmetric tensor is called
{\bf hypersurface data}.

Given hypersurface data, we define
a torsion-free connection $\nablab$ on $\N$ as the connection with connection
coefficient symbols given by (\ref{GambProp}). We also define the tensors
$\Kn_{ab}$, $\varphi_a$ and $\Psi^{b}_{\,\,\,a}$ by
the expressions (\ref{KnProp})-(\ref{PsiProp}).
\end{definition}

As before, a notion of ``embedding'' for hypersurface data becomes
necessary in order to link the hypersurface data
with the ambient spacetime expressions (\ref{covder1})-(\ref{covder2}).

\begin{definition}[Embedding of hypersurface data]
Let $\{ \N,\gamma,\ellc,\ll,{\bm Y} \}$ be hypersurface data. We will say
that this data is embedded in a spacetime $(\M,\gM)$ if there
exists an embedding $\Phi: \N \longrightarrow \M$ and a choice
of rigging $\ell$ such that, with $\bmell = g ( \ell,\cdot)$,
\begin{eqnarray*}
\Phi^{\star} (\gM) = \gamma, \quad \Phi^{\star} \left ( \gM(\ell, \cdot) \right ) =
\ellc,
\quad   \gM(\ell,\ell)  = \ll, \quad
\frac{1}{2} \Phi^{\star} \left ( {\pounds}_{\ell\,\,} \gM \right ) = {\bf Y}.
\end{eqnarray*}
\end{definition}

\vs
\noindent \noindent {\bf Remark \arabic{remark}. \addtocounter{remark}{1}}
The last formula of the definition requires an extension of the rigging 
$\ell$ off $\Phi(\Sigma)$. The expression is, however, independent of this
extension. Note also that, for embedded hypersurface data, the tensor
$Y_{ab}$ corresponds to the symmetric part 
of the tensor ${\cal H}_{ab}$ introduced, and extensively used, 
in \cite{MarsSenovilla1993} (see also \cite{MarsSenovillaVera2007}).

\vs

For this definition to make sense it is necessary that
the covariant derivative
$\nablab$ and the tensors $\Kn_{ab}$, $\varphi_{a}$, $\Psi^{b}_{\,\,\,b}$
defined by Proposition \ref{solving}
coincide with the corresponding
tensors  defined via (\ref{GambAmb}), (\ref{KnAmb}), (\ref{varphiAmb}),
(\ref{PsiAmb}) in terms of the ambient spacetime geometry. This is taken care
of in the following lemma.
\begin{lemma}
Let $\{\N,\gamma,\ellc,\ll,{\bm Y} \}$ be
hypersurface data 
and $\{ \Gamb^{c}_{ab}, \Kn_{ab}, \varphi_{a}, \Psi^{b}_{\,\,\,a} \}$
be defined by (\ref{GambProp})-(\ref{PsiProp}). Assume that this hypersurface
data is embedded with embedding  $\Phi$ and rigging vector $\ell$
and let $e_a \defi \Phi_{\star} (\hat{e}_a)$
where $\{ \hat{e}_a \}$ be a basis of $T \N$. Then,
the field equations (\ref{covder1}) and (\ref{covder2}) are satisfied.
\end{lemma}

\noindent {\it Proof.} Define (a priori new) tensors $\tilde{\Kn}_{ab} =
\tilde{\Kn}_{ba}$, $\tilde{\varphi}_{a}$, $\tilde{\Psi}^{b}_{\,\,\,a}$
and connection coefficients $\tilde{\Gamb}^{c}_{ab}$ by 
the decompositions $\nabla_{e_a} e_b  =   - \tilde{\Kn}_{ab}  \ell + 
\tilde{\Gamb}^{c}_{ba}
e_c$ and $\nabla_{e_a} \ell = 
\tilde{\varphi}_a  \ell + \tilde{\Psi}^b_{\,\,\,a} e_b$.
Then equations (\ref{comp1})-(\ref{comp3}) are satisfied
by this fields and hence, by Proposition \ref{solving}, there
exists a tensor $\tilde{Y}_{ab}$ such that (\ref{GambProp})-(\ref{PsiProp})
are satisfied with the substitutions
$\{ \Gamb^c_{ab}, \Kn_{ab}, \varphi_{a}, \Psi^{b}_{\,\,\,a}, Y_{ab} \} 
\longrightarrow
\{ \tilde{\Gamb}^c_{ab}, \tilde{\Kn}_{ab}, \tilde{\varphi}_{a}, 
\tilde{\Psi}^{b}_{\,\,\,a}, Y_{ab} \}$. Thus, to prove the Lemma
we only need to make sure that $Y_{ab} = \tilde{Y}_{ab}$. Now,
from the definition of embedded hypersurface data
\begin{align*}
2 Y_{ab} & = e_a^{\, \mu} e_b^{\, \nu}
\left ( \nabla_{\mu} \ell_{\nu}
+ \nabla_{\nu} \ell_{\mu} \right ) = 
 (\nabla_{e_a} {\bmell}) (e_b) 
+ (\nabla_{e_b} {\bmell}) (e_a) = \tilde{\varphi}_a \ell_b 
+ \tilde{\Psi}^{c}_{\,\,\,a} \gamma_{cb} 
+ \tilde{\varphi}_b \ell_a 
+ \tilde{\Psi}^{c}_{\,\,\,b} \gamma_{ca}  = \\
& = 
 \frac{}{} \widetilde{\nablab}_{a} \ell_b + \ll \tilde{\Kn}_{ab}
+ \widetilde{\nablab}_{b} \ell_a + \ll \tilde{\Kn}_{ab} = 
\widetilde{\nablab}_{a} \ell_b  + \widetilde{\nablab}_{b} \ell_a  + 2
\ll \tilde{\Kn}_{ab}  = 2 \tilde{Y}_{ab}
 \end{align*}
where in the fourth equality we have used the tilded version of (\ref{comp2})
($\widetilde{\nablab}$ is the covariant derivative with connection 
$\widetilde{\Gamb}^{c}_{ab}$)  and the last equality follows
from (the tilded version of) Proposition \ref{solving} . \ep

\vs

Given hypersurface metric data, we have defined $P^{ab}$, $n^c$ and $\nn$
as the solutions of equations (\ref{EqP1})-(\ref{EqP4}). If
the data is supplemented with $Y_{ab}$ to yield hypersurface data, then
equations (\ref{comp1})-(\ref{comp3}) are identically satisfied. 
It is clear that the fields $P^{ab}$, $n^c$ and $\nn$  will also
satisfy appropriate field equations. If the hypersurface
is embedded in a spacetime, the equations are easily derived by a calculation
similar to derivation above leading to (\ref{comp1})-(\ref{comp3}).
However, as we want
to work at the data level alone, we need to argue directly with
the expressions on $\N$. In the following proposition we obtain 
a number of identities that will immediately imply the equations we are 
looking for.
We note that the definitions of $\A_{abc}$, $\B_{ab}$, $\C_{a}$
in the proposition come directly from (\ref{comp1})- (\ref{comp3})
while the definitions of $\D_{abc}$, $\E_{ab}$, $\F_{a}$ are motivated
by the spacetime calculation indicated above.

\begin{proposition}
\label{invEqs}
Let $n^a$, 
$\ell_a$, $\gamma_{ab} = \gamma_{ba}$, $\ll$, $P^{ab} = P^{ba}$,
$n^a$, $\nn$ be arbitrary
$C^1$ tensor fields on a manifold $\N$ endowed with a
connection $\nablab$. Let $\Kn_{ab} = \Kn_{ba}$, $\Psi^{a}_{\,\,\,b}$
and $\varphi_a$ arbitrary $C^0$ tensor fields on $\N$. Define
\begin{eqnarray}
 \A_{abc} & \defi & \nablab_a \gamma_{bc} + \ell_b \Kn_{ac} + \ell_c \Kn_{ab},
\nonumber  \\
 \B_{ab} & \defi & \nablab_a \ell_{b} - \varphi_a \ell_b + \ll \Kn_{ab}
- \Psi^{c}_{\,\,\,a} \gamma_{bc} \nonumber \\
 \C_a & \defi &  - \frac{1}{2} \nablab_a \ll + \ll \varphi_a + \Psi^{b}_{\,\,\,a}
\ell_b \nonumber \\
 \D_a^{\,\,\,bc} & \defi & \nablab_a P^{bc} + n^b \Psi^{c}_{\,\,\,a}
+ n^c \Psi^{b}_{\,\,\,a}, \label{defall} \\
 \E_a^{\,\,\,b} & \defi & 
\nablab_a n^b  + \varphi_a n^b- P^{bc} \Kn_{ac} 
+  \nn \Psi^{b}_{\,\,\,a}, \nonumber \\
 \F_a & \defi & - \frac{1}{2} \nablab_a  \nn - \nn \varphi_a 
+ \Kn_{ab} n^b. 
\end{eqnarray}
and
\begin{align}
&  q \defi  n^a \ell_a - 1 + \nn \ll, \quad  
& & z_b  \defi  \gamma_{bc} n^c + \nn \ell_b, \nonumber \\
& w^a  \defi  P^{ab} \ell_b + \ll n^a, \quad
& & 
 S^a_{\,\,\,b}  \defi  P^{ac} \gamma_{cb} + n^a \ell_b - \delta^{a}_{b}.
\label{defqs}
\end{align}
Then, the following identities hold
\begin{align}
 \E_{a}^{\,\,\,b} \ell_b - 2 \ll \F_{a} - 2 \nn \C_a  & =  
\nablab_a q - \Kn_{ac} w^c -  \Psi^{c}_{\,\,\,a} z_c, \nonumber \\
 \A_{abc} n^c +  \E_a^{\,\,\,c} \gamma_{bc}
 - 2 \F_a \ell_b + \nn \B_{ab}
& = \nablab_a z_b + \Kn_{ac} \left ( q \delta^c_{b} - S^c_b
\right ) + \varphi_a z_b, \label{identities}  \\
 \D_{a}^{\,\,\,bc} \ell_c + P^{bc} \B_{ac} - 2 n^b \C_a 
+ \ll \E_a^{\,\,\,b} & = \nablab_a w^b + \Psi^{c}_{\,\,\,a}
\left ( q \delta^{b}_{c} - S^b_c \right ) - \varphi_a w^b, \nonumber \\
\D_a^{\,\,\,bd} \gamma_{dc} + P^{bd} \A_{adc} + \E_a^{\,\,\,b} \ell_c
+ n^b \B_{ac} &  = \nablab_{a} S^b_c + \Psi^b_{\,\,\,a} z_c + \Kn_{ac} w^b.
\nonumber 
\end{align}
\end{proposition}

\vs 

\noindent {\it Proof.} The proof is by straightforward (and somewhat long)
calculation.
We give the proof explicitly for the first identity. From the
definition of $q$, its derivative reads
\begin{align*}
\nablab_a q & = \ell_b \nablab_a n^b + n^b \nablab_a \ell_b 
+ \ll \nablab_a \nn + \nn \nablab_a \ll \\
& = 
\left ( \E_a^{\,\,\,b} + P^{bc} \Kn_{ac} - \nn \Psi^{b}_{\,\,\,a}
\right ) \ell_b + n^b \left ( \B_{ab} - \ll \Kn_{ab}
+ \Psi^{c}_{\,\,\,a} \gamma_{bc} \right )
+ 2 \ll \left ( \Kn_{ac} n^c - \F_{a} \right )  + 2 \nn  \left (
\Psi^{c}_{\,\,\,a} \ell_c - \C_{a} \right )  \\
& = \E_{a}^{\,\,\,b} \ell_b - 2 \ll \F_{a} - 2 \nn \C_a 
+ \Kn_{ac} w^c +  \Psi^{c}_{\,\,\,a} z_c,
\end{align*}
where in the second equality we have used the definitions 
(\ref{defall}) and in the third we have used
the definitions (\ref{defqs}). The rest of expressions  are more involved
but can be proved similarly. \ep

\vs 
The following corollary determines the equations that $P^{ab}$,
$n^b$ and $\nn$ satisfy for hypersurface data.
\begin{corollary}
Let $\{\N,\gamma_{ab},\ell_a,\ll,Y_{ab}\}$ be hypersurface data. Then, the
following equations hold
\begin{align}
\nablab_a P^{bc} + n^b \Psi^{c}_{\,\,\,a}
+ n^c \Psi^{b}_{\,\,\,a} = 0, \label{eqsinv1} \\
\nablab_a n^b  + \varphi_a n^b- P^{bc} \Kn_{ac} 
+  \nn \Psi^{b}_{\,\,\,a} = 0, \label{eqsinv2}\\
- \frac{1}{2} \nablab_a  \nn - \nn \varphi_a 
+ \Kn_{ab} n^b = 0. \label{eqsinv3}
\end{align}
\end{corollary}
\noindent {\it Proof.} 
Hypersurface data satisfies $q=z_a=w^a=S^a_b=0$
and also $\A_{abc} = \B_{ab} = \C_a =0$. Identities (\ref{identities})
then become
\begin{eqnarray*}
 \E_{a}^{\,\,\,b} \ell_b - 2 \ll \F_{a}    = 0, \quad \quad
\E_a^{\,\,\,c} \gamma_{bc}  - 2 \F_a \ell_b 
 = 0,  \\
 \D_{a}^{\,\,\,bc} \ell_c + \ll \E_a^{\,\,\,b}  = 0, \quad \quad 
\D_a^{\,\,\,bd} \gamma_{dc} + \E_a^{\,\,\,b} \ell_c  = 
0.
\nonumber 
\end{eqnarray*}
Applying Lemma \ref{raising}  to the first two (with $\E_{a}^{\,\,\,b}
 \rightarrow  V^b$) yield the compatibility equation
\begin{equation*}
0 = \F_{a} \left ( n^b l_b + \nn \ll\right ) = \F_a
\end{equation*}
and hence $\E_a^{\,\,\,b}=0$ also. Applying now Lemma \ref{raising}
to the second two gives $\D_{a}^{\,\,\,bc} = 0$. 
\ep

\vs

Proposition  \ref{invEqs} has a second consequence on 
the relationship between $\{P^{ab},n^b,\nn\}$
and $\{\gamma_{ab},\ell_a,\ll\}$, whenever the
field equations (\ref{eqsinv1})-(\ref{eqsinv3}) hold
together with the field
equations (\ref{comp1})-(\ref{comp3}):
\begin{corollary}
Let $\gamma_{ab} = \gamma_{ba}$, $\ell_a$, $\ll$,
$P^{ab} = P^{ba}$, $n^b$, $\nn$ be $C^1$ fields on an $m$-dimensional
connected manifold $(\N,\nablab)$ satisfying
the field equations (\ref{comp1})-(\ref{comp3}) and 
(\ref{eqsinv1})-(\ref{eqsinv3}). If the two $(m+1)$-matrices
\begin{equation*}
\begin{bmatrix}
\gamma_{ab} & \ell_a \\
\ell_b & \ll 
\end{bmatrix},  \quad \quad
\begin{bmatrix}
P^{ab} & n^a \\
n^b & \nn 
\end{bmatrix} 
\end{equation*}
are inverses from each other at one point $p\in \N$, then they are inverses
of each other at every point in $\N$.
\end{corollary}
\noindent {\it Proof.} Identities (\ref{identities}) with 
$\A_{abc} = \B_{ab} = \C_{a} = \D_{a}^{\,\,\,bc} = \E_a^{\,\,\,b} = \F_{a} = 0$
become a set of linear PDE for $\{ q,z_a, w^a, S^{a}_{b}\}$ written in 
normal form. Thus, if $q = z_a = w^a = S^a_b=0$ at one point, then
they vanish everywhere on the (connected) manifold $\N$. \ep

\section{Gauge transformations.}
\label{gaugeTrans}

Up to now we have kept the rigging fixed. However, as already
said, the rigging is highly non-unique. Thus, there must exist a set
of transformations that keep the field equations invariant and which give
essentially the same hypersurface data. This section is devoted to
this issue.

Let us for the moment consider a hypersurface embedded in a spacetime. We
will find how does the hypersurface data transform under an arbitrary change of rigging. 
Then, we will promote this transformation to  gauge freedom in the
hypersurface data and we will prove that the field equations are invariant
under a gauge transformation.

Since a rigging is, by definition, transverse to the hypersurface 
any two riggings $\ell$ and $\ell'$ are
related by \cite{MarsSenovilla1993}
\begin{eqnarray}
\ell' \eqN u \left (  \ell  +  V \right),
\label{gaugel}
\end{eqnarray}
where $u$ is nowhere zero and $V$ is an arbitrary 
vector field along $\N$ and tangent to $\N$ everywhere. $V$
can be decomposed in the basis $\{ e_a \}$ as
$V = V^a e_a$. The vector $\hat{V} \defi V^a \hat{e}_a$
is therefore a vector field of $\N$.

First of all we note that the first fundamental
form $\gamma_{ab}$ is independent of the choice of
rigging and hence $\gamma'_{ab} = \gamma_{ab}$ (objets attached to the rigging
$\ell'$ will carry a prime). 
Multiplying (\ref{gaugel}) by $e_a$ we obtain
\begin{eqnarray*}
\ell'_a = u \left ( \ell_a + V^b \gamma_{ab} \right ).
\end{eqnarray*} 
and squaring $\ell'$  we find
\begin{eqnarray}
\ell'{}^2= \gM( \ell',\ell' ) 
= u^2 \left ( \ll + 2 V^a \ell_a 
+ V^a V^b \gamma_{ab} \right ). \label{gaugell}
\end{eqnarray}
It only remains to determine how does the tensor ${\bm Y}$ change
under a gauge transformation. Since we are assuming
the data to be embedded, we can use ${\bm Y} = \frac{1}{2} \Phi^{\star} 
\left ( \pounds_{\ell} \gM \right )$. We can determine 
${\bm Y}^{\prime}$ as follows. Let $\tilde{u}$ denote
any smooth extension of $u$ off $\N$. Then
\begin{align*}
{\bm Y}^{\prime} & = 
\frac{1}{2} \Phi^{\star} \left ( \pounds_{\ell^{\prime}} \gM \right ) = 
\frac{1}{2} \Phi^{\star} \left ( \pounds_{\tilde{u} (\ell + V)} \gM \right ) = 
\frac{1}{2} \Phi^{\star} \left ( \tilde{u} \pounds_{\ell} \gM 
+ d \tilde{u} \otimes \bmell + \bmell \otimes d \tilde{u} 
+ \pounds_{\tilde{u} V} \gM \right ) =  \\
& = u \bm{Y} + \frac{1}{2} \left ( du \otimes \ellc + \ellc \otimes
du \right ) + \frac{1}{2} \pounds_{u \hat{V}} \gamma,
\end{align*}
where we used the well-known properties 
$\Phi^{\star} (d \tilde{u}) = d (\Phi^{\star}  (\tilde{u} ))$
and $\Phi^{\star} ( \pounds_{ \Phi_{\star} (\hat{V})} \gM ) = \pounds_{\hat{V}} \left ( \Phi^{\star} (\gM )\right )$.

The transformations above only involve a scalar function $u$ on $\N$
and a vector field $\hat{V}$ on $\N$. We can therefore 
put forward a definition of {\it gauge transformation}
for hypersurface data.

\begin{definition}
\label{gauge}
Let $\{\gamma,\ellc,\ll, {\bm Y} \}$ be hypersurface
data. Let $u: \N \rightarrow \mathbb{R}$ be a smooth scalar and
$V \in \Gamma ( T \N )$ a smooth vector field in $\N$.
The {\bf gauge transformed} hypersurface data with gauge fields $(u,\hat{V})$ 
is defined as (in any basis $\{ \hat{e}_a \}$ and with $\hat{V} =
V^a \hat{e}_a$)
\begin{align}
\gamma^{\prime}_{ab} & = \gamma_{ab}, \quad \ell^{\prime}_a =
u \left (\ell_a + V^b \gamma_{ab} \right ), \quad
\llprime = u^2 \left ( \ll + 2 V^a \ell_a + V^a V^b \gamma_{ab} \right ), 
\nonumber \\
Y^{\prime}_{ab} & = u Y_{ab} + \frac{1}{2} \left ( \ell_a \partiald_b u 
+ \ell_b \partiald_{a} u  \right ) +  \frac{1}{2} \pounds_{u \hat{V} } \gamma_{ab}.
\label{gaugetrans}
\end{align}
\end{definition}

The following Lemma shows how the derived fields
$\{ P^{ab}, n^a, \nn\}$ and
$\{ \Gamb^{c}_{ab}, \Kn_{ab}, \varphi_{a}, \Psi^{b}_{\,\,\,a} \}$  
change  under a gauge transformation (cf. \cite{MarsSenovilla1993} and
the Remark after the proof). 
\begin{lemma}
\label{gaugetransformation}
Let $(u,V^a)$ be gauge fields. The fields 
$(P^{\prime \, ab}, n^{\prime \, a}, \nnprime)$
corresponding to the gauge
transformed data read
\begin{eqnarray}
 P^{\prime \, ab}  =   P^{ab} + V^a V^b \nn - V^a n^b - V^b n^a, \quad
 n^{\prime \, a}  =   u^{-1} (  n^a - V^a \nn ),
\quad \nnprime   =  u^{-2} \nn. \label{Pnprime}
\end{eqnarray}
Moreover, the connection $\Gamb^{c}_{ab}$
and the tensor fields $\Kn_{ab}, \varphi_{a}, \Psi^{b}_{\,\,\,a}$ 
transform according to 
\begin{align}
\Gamb^{\prime} {}^{c}_{ab} & = \Gamb^{c}_{ab} + V^c K_{ab}, \label{Gambprime} \\
\Kn^{\prime}_{ab} & =  u^{-1} \Kn_{ab},
 \label{Knprime} \\
\varphi^{\prime}_{a} & = \varphi_a + u^{-1} \partiald_a u - \Kn_{ab} V^b,
\label{varphiprime} \\
\Psi^{\prime} {}^{b}_{\,\,\,a} & = 
u \left ( \Psi^b_{\,\,\,a} + \nablab_a V^b - \varphi_a V^b
+ \Kn_{ad} V^d V^b \right ).
    \label{Psiprime}
\end{align}
\end{lemma}
\vs
\noindent {\it Proof.} A straightforward calculation shows that, with 
the expressions (\ref{Pnprime}), the primed
version of (\ref{EqP1})-(\ref{EqP4}) is satisfied.
Uniqueness of solutions of these equations
implies the first part of the Lemma.
Next we prove (\ref{Knprime}). First notice the following simple identity
for the Lie derivative (a version of which was in fact 
already used above).
\begin{eqnarray}
A \pounds_{\hat{X}} \gamma_{ab} 
=  \pounds_{A \hat{X}} \gamma_{ab} - \gamma_{ac} X^c \partial_b A
- \gamma_{bc} X^c \partial_a A.
\label{idenA}
\end{eqnarray}
We now calculate $\nnprime {\bm Y}^{\prime}$
\begin{align*}
\nnprime Y^{\prime}_{ab} & = 
\frac{\nn}{u} Y_{ab} + \frac{\nnprime}{2}
\left ( \ell_a \partial_b u + \ell_b \partial_a u \right ) + 
\frac{\nnprime}{2} \pounds_{u \hat{V}} \gamma_{ab} \\
& = 
\frac{\nn}{u} Y_{ab} + \frac{\nnprime}{2}
\left ( \ell_a \partial_b u + \ell_b \partial_a u \right ) + 
\frac{1}{2} \pounds_{\frac{\nn}{u} \hat{V}} \gamma_{ab} 
- \frac{1}{2} \left (u \gamma_{ac} V^c \partial_b (\nnprime)
+ u\gamma_{bc} V^c \partial_a (\nnprime) \right ) \\
& =
\frac{\nn}{u} Y_{ab} + \frac{\nnprime}{2}
\left ( \ell_a \partial_b u + \ell_b \partial_a u \right ) + 
\frac{1}{2} \pounds_{\frac{\nn}{u} \hat{V}} \gamma_{ab} 
+ \frac{1}{2} \left ( u \ell_a - \ell^{\prime}_a \right )  \partial_b (\nnprime)
+ \frac{1}{2} \left ( u \ell_b - \ell^{\prime}_b \right )  \partial_a (\nnprime)
\end{align*}
where in the second equality we used 
(\ref{idenA}) with $\hat{X} 
\rightarrow u \hat{V}$ and $A \rightarrow \nnprime$,
and in the third equality we used the transformation law for $\ell_{a}$.
We now insert this into the primed version of (\ref{KnProp}). This yields
\begin{align*}
\Kn^{\prime}_{ab} & = 
\frac{\nn}{u} Y_{ab} + \frac{\nnprime}{2}
\left ( \ell_a \partial_b u + \ell_b \partial_a u \right ) 
+ \frac{1}{2} \left ( u \ell_a \partial_b (\nnprime)
+ u \ell_b \partial_a (\nnprime) \right )
+ \frac{1}{2} \pounds_{\hat{n}^{\prime} + \frac{\nn}{u} \hat{V}} \gamma_{ab}  \nonumber  \\
& =
\frac{1}{u}
\left ( \nn Y_{ab} + \frac{1}{2} \left ( \ell_a \partial_b \nn
+ \ell_b \partial_a \nn \right ) + \frac{1}{2} \pounds_{\hat{n}} \gamma_{ab}
\right )  - \frac{1}{2 u^2 }  \left ( \nn \ell_a - \gamma_{ac} n^c
\right ) \partial_b u  - \frac{1}{2 u^2}
\left ( \nn \ell_b - \gamma_{bc} n^b \right ) \partial_a u \\
& = \frac{1}{u} \Kn_{ab}, 
\end{align*}
where in the second equality we have used $\hat{n}^{\prime} 
+ \frac{\nn}{u} \hat{V} = u^{-1} \hat{n}$ and
then the identity (\ref{idenA}) with $\hat{X} 
\rightarrow u^{-1} \hat{n}$ and $A \rightarrow u$
and in the last equality we have used (\ref{EqP3}) and (\ref{KnProp}).
This proves (\ref{Knprime}).

Next we address (\ref{Gambprime}). $\Gamb^{\prime}{}^{c}_{ab}$ is uniquely
defined as the unique solution of the primed versions of (\ref{ExpGambgam})
and (\ref{ExpGambell}). Subtracting (\ref{ExpGambgam}) and its primed
version and using the invariance of $\gamma_{ab}$ and the
transformation law for $\ell_{a}$ and $\Kn_{ab}$ it follows
\begin{eqnarray}
\left 
( \Gamb^{\prime}{}^{d}_{ab} - \Gamb^{d}_{ab} \right ) 
\gamma_{cd} = 
V^d \gamma_{cd} K_{ab} \label{difgam}
\end{eqnarray}
If we also determine 
$( \Gamb^{\prime}{}^{d}_{ab} - \Gamb^{d}_{ab}  ) \ell_{d}$ we will be able to 
solve for $\Gamb^{\prime} {}^{d}_{ab}$. To that aim, let us start by
deriving the following identity
\begin{align}
\label{idenderlY}
& \frac{1}{2} \left ( \frac{}{} \partial_a \ell^{\prime}_b + \partial_b 
\ell^{\prime}_a
\right ) - Y^{\prime}_{ab} = \nonumber \\
& \hspace{7mm} = 
\frac{1}{2} \left ( \frac{}{} 
\partial_a \left [ u ( \ell_b +  \gamma_{bc} V^c )\right  ] + 
\partial_b \left [ u ( \ell_a +  \gamma_{ac} V^c ) \right ] \right ) -
u Y_{ab} 
- \frac{1}{2} \left ( \ell_a \partial_b u + \ell_b
\partial_a u \right ) - \frac{1}{2} \pounds_{u \hat{V}} \gamma_{ab} 
\nonumber \\
& \hspace{7mm} =
u \left ( \frac{1}{2} \left ( \partial_a \ell_b + 
\partial_a \ell_b \right ) - Y_{ab} \right )
+ \frac{1}{2} \left [ \partial_{a} \left ( u \gamma_{bc} V^c \right ) 
+ \partial_b \left ( u \gamma_{ac} V^c \right )  - \pounds_{u \hat{V}} \gamma_{ab}
\right ] \nonumber \\
& \hspace{7mm} = 
u \left ( \frac{1}{2} \left ( \partial_a \ell_b + 
\partial_a \ell_b \right ) - Y_{ab} \right )
+ \frac{u}{2} \left ( \partial_a \gamma_{bc} + 
\partial_{b} \gamma_{ac} - \partial_c \gamma_{ab} \right ) V^c.
\end{align}
Now we evaluate 
\begin{align*}
\left ( \Gamb^{\prime}{}^{d}_{ab} -
\Gamb^{d}_{ab} \right ) \ell_d  = \,\, & 
\frac{1}{u} \Gamb^{\prime} {}^{d}_{ab} \ell^{\prime}_{d} 
- \Gamb^{d}_{ab} \ell_{d} - 
\Gamb^{\prime} {}^{d}_{ab} \gamma_{dc} V^c \\
 = \, \, & 
\frac{1}{u} 
\left ( 
\frac{1}{2} \left ( \frac{}{} \partial_a \ell^{\prime}_b +
\partial_b \ell^{\prime}_a \right ) - Y^{\prime}_{ab} 
+ \frac{1}{u} \llprime K_{ab} \right )  - \left ( 
\frac{1}{2} \left ( \frac{}{} \partial_a \ell_b +
\partial_b \ell_a \right ) - Y_{ab} 
+  \ll K_{ab} \right ) \\
& - \frac{1}{2} \left ( \partial_{a} \gamma_{bc} +
\partial_{b} \gamma_{ac} - \partial_c \gamma_{ab}
\right ) V^c 
- \left ( \ell_c + \gamma_{cd} V^d \right ) V^c K_{ab} \\
 = \,\, &  V^c \ell_c  K_{ab}
\end{align*}
where in the second equality we have used the primed version
of (\ref{ExpGambell}), equation (\ref{ExpGambell}) itself
and the primed version of (\ref{ExpGambgam})
and in the last equality we used the identity
(\ref{idenderlY}) and the transformation law for $\ll$. It is now
immediate from this expression and (\ref{difgam})
to conclude that (\ref{Gambprime}) holds as a consequence of Lemma
\ref{raising}.

In order to prove the remaining transformations, we first
establish the following transformation ($\nablab^{\prime}$
is the covariant derivative with connection symbols $\Gamb^{\prime} {}^{c}_{ab}$) 
\begin{align}
\nablab^{\prime}_a \ell^{\prime}_b + \llprime \Kn'_{ab} & =
\nablab_{a} \ell^{\prime}_b - \left ( \Gamb^{\prime}{}^{c}_{ab} - 
\Gamb^{c}_{ab} \right ) \ell^{\prime}_c + \llprime \Kn'_{ab} \nonumber \\
& =
\nablab_a \left ( u \left [ \ell_{b} + \gamma_{bc} V^c \right ]
\right ) - u V^c \Kn_{ab}  \left ( \ell_c + \gamma_{cd} V^d \right )
+ u^{-1} \llprime  \Kn_{ab} \nonumber \\
& =
u \left ( \nablab_{a} \ell_b + \ll \Kn_{ab} \right )
+ \left ( \ell_b + \gamma_{bc} V^c \right ) \nablab_a u 
- u \Kn_{ac} V^c \ell_b + 
u \gamma_{bc} \nablab_a V^c. \label{nablaell}
\end{align}
where in the second equality we used the 
gauge transformation law for $\ell_c$, $\Gamb^{c}_{ab}$ and $\Kn_{ab}$
and in the third one the field equation (\ref{comp1}) and
the transformation law for $\ll$ was used. Proposition \ref{solving}
shows that $\varphi_a$ can be written in the form
\begin{eqnarray}
\varphi_a = n^b \left ( \nablab_a \ell_b + \ll \Kn_{ab} \right ) + 
\frac{\nn}{2} \nablab_a \ll.
\label{varphi_two}
\end{eqnarray}  
The transformation law (\ref{varphiprime}) for $\varphi_a$
follows by a direct substitution of $n^{\prime \, b}$,
$\nnprime$, $\llprime$ and (\ref{nablaell}) in the primed version
of (\ref{varphi_two}).

The proof of (\ref{Psiprime}) is the most involved one. One option would
be  brute force calculation from (\ref{PsiProp}). However, the
gauge transformation of $P^{ab}$ is long and the calculation becomes cumbersome. 
Instead, we subtract the primed version of (\ref{comp2}) and 
$u$ times (\ref{comp2}) itself. This yields, after using (\ref{nablaell})
and the transformation law for $\varphi_a$,
\begin{eqnarray}
\gamma_{bc} \left ( 
\Psi^{\prime}{}^c_{\,\,\,a} -
 u \left ( \Psi^c_{\,\,\,a} + \nablab_a V^c - \varphi_a V^c
+ \Kn_{ad} V^d V^c \right ) \right ) = 0.
\label{DifPsigam}
\end{eqnarray}
This still fails short of proving (\ref{Psiprime}) because
$\gamma_{bc}$ need not be invertible. To complete the argument, we 
compute the primed version of (\ref{comp3}) subtracting $u^2$ times
(\ref{comp3}) and adding $u^2$ times (\ref{comp2}) contracted with $V^b$.
A not-long calculation which uses (\ref{DifPsigam}) yields
\begin{eqnarray}
u \ell_c \left (
\Psi^{\prime}{}^c_{\,\,\,a} -
 u \left ( \Psi^c_{\,\,\,a} + \nablab_a V^c - \varphi_a V^c
+ \Kn_{ad} V^d V^c \right ) \right ) = 0.
\label{DifPsiell}
\end{eqnarray}
Lemma \ref{raising} applied to (\ref{DifPsigam}) and (\ref{DifPsiell})
establishes (\ref{Psiprime}). \ep

\vs 

\noindent {\bf  Remark \arabic{remark}. \addtocounter{remark}{1}} The gauge transformation law obtained in this
proposition would have been much easier to prove assuming that the
hypersurface data is embedded. Indeed, in such circumstances, the 
spacetime definition of the connection, the second fundamental form
and the tensors $\Psi^{b}_{\,\,\,a}$, $\varphi_{a}$ imply very
easily (\ref{Gambprime})-(\ref{Psiprime}), see \cite{MarsSenovilla1993}.
However, our definition of gauge transformation
is directly at the hypersurface data level, and hence a hypersurface proof
as the one above becomes necessary.

\vs From  the proof of the preceeding lemma, we can infer that
in order to find gauge transformation laws of contravariant tensors,
it may be often convenient to study the transformation law of the one-form
obtained after contraction with $\gamma_{ab}$ and the scalar obtained
by contraction with $\ell_b$. The following proposition formalizes
this observation.

\begin{proposition}
\label{TransContractions}
Let $Q^a = Q^a(\gamma_{bd}, \ell_b, \ll, Y_{bd})$ be a vector depending on the
hypersurface data and define
$H_c \defi Q^a \gamma_{ac}$, $R \defi Q^a \ell_a$, all of them viewed
as functions of the hypersurface data. Fix gauge
fields $(u,V^a)$ and let
%$H^{\prime}_c \defi H_c (\gamma_{bd}, \ell^{\prime}_b, \llprime, Y^{\prime}_{bd})$
%and $R^{\prime} \defi R (\gamma_{bd}, \ell^{\prime}_b, \llprime, Y^{\prime}_{bd})$.
$\hat{H}_c$, $\hat{R}$ be the functions of hypersurface data and gauge field
defined by
\begin{eqnarray*}
\hat{H}_c (\gamma_{bd}, \ell_b, \ll, Y_{bd}, u, V^d ) \defi H_{c} (\gamma_{bd}, \ell^{\prime}_b, \llprime, Y^{\prime}_{bd}),
\quad \quad
\hat{R} (\gamma_{bd}, \ell_b, \ll, Y_{bd}, u, V^d )
\defi R (\gamma_{bd}, \ell^{\prime}_b, \llprime, Y^{\prime}_{bd}),
\end{eqnarray*}
where in the right-hand side we substitute
 the explicit expressions (\ref{gaugetrans}). 
Then, the following two statements are equivalent.
\begin{itemize}
\item[(i)] The functions
$\hat{Q}^a (\gamma_{bd}, \ell_b, \ll, Y_{bd}, u, V^d)$ satisfy
the identities
\begin{align}
& \hat{H}_c =  \hat{Q}^a \gamma_{ac}, \label{id1} \\
& \hat{R} =  u \left ( \hat{Q}^a \ell_a + V^a \hat{H}_a \right ).
\label{id2} 
\end{align}
\item[(ii)] The functions
$\hat{Q}^a (\gamma_{bd}, \ell_b, \ll, Y_{bd}, u, V^d)$ satisfy
$\hat{Q}^a (\gamma_{bd}, \ell_b, \ll, Y_{bd}, u, V^d) =
Q^a (\gamma_{bd}, \ell^{\prime}_b, \llprime, Y^{\prime}_{bd})$, where in the
right-hand side the explicit expressions 
(\ref{gaugetrans}) are substituted.
\end{itemize}
\end{proposition}

\vs

\noindent {\it Proof.} Assume that (ii) holds. Then
\begin{align*}
& \hat{H}_c =   H_c (\gamma_{bd}, \ell^{\prime}_b, \llprime, Y^{\prime}_{bd})
= Q^a (\gamma_{bd}, \ell^{\prime}_b, \llprime, Y^{\prime}_{bd}) \gamma_{ac} =
\hat{Q}^a \gamma_{ac}, \\
& \hat{R} =  R (\gamma_{bd}, \ell^{\prime}_b, \llprime, Y^{\prime}_{bd})
= 
Q^a (\gamma_{bd}, \ell^{\prime}_b, \llprime, Y^{\prime}_{bd}) \ell^{\prime}_{a} =
u \left ( 
Q^a (\gamma_{bd}, \ell^{\prime}_b, \llprime, Y^{\prime}_{bd}) 
\ell_a + V^c Q^a (\gamma_{bd}, \ell^{\prime}_b, \llprime, Y^{\prime}_{bd}) 
\gamma_{ac} \right ) = \\
& =  u \left ( \hat{Q}^a \ell_a 
+ V^a \hat{H}_{a} \right ), 
\end{align*}
which establishes the validity of (i). Conversely,
assume that (i) holds. 
Define $\tilde{Q}^a (\gamma_{bd}, \ell_b, \ll, Y_{bd}, u, V^d) =
Q^a (\gamma_{bd}, \ell^{\prime}_b, \llprime, Y^{\prime}_{bd})$, where in the
right hand side the explicit expressions 
(\ref{gaugetrans}) are substituted. By the calculation above, this 
function satisfies
\begin{align*}
& \hat{H}_c =  \tilde{Q}^a \gamma_{ac}, \\
& \hat{R} =  u \left ( \tilde{Q}^a \ell_a + V^c \hat{H}_c \right ).
\end{align*}
Subtracting this to (\ref{id1}) and (\ref{id2}) yields
$ ( \hat{Q}^a - \tilde{Q}^a ) \gamma_{ac} =0$ and 
$ u ( \hat{Q}^a - \tilde{Q}^a ) \ell_a =0$. Lemma \ref{raising} implies
$ \hat{Q}^a - \tilde{Q}^a =0$ which proves (ii). \ep

\vs

\noindent {\bf  Remark \arabic{remark}. \addtocounter{remark}{1}} Obviously, this result can be applied to tensors
of arbitrary rank as long as one of its indices is contravariant and all
the operations and expressions involved in the lemma refer to this index,
leaving all the rest of indices untouched.

\vs 

As an example of the usefulness of this lemma, we obtain the
transformation law of $(\nn P^{ab} - n^a n^b)$
and of $P^{ab} n^c - P^{ac} n^b$, which will be needed later.
\begin{lemma}
\label{TranPnn}
Under a gauge transformation with gauge fields $(u,V^a)$, the tensors
$\nn P^{ab} - n^a n^b$  and $P^{ab} n^c - P^{ac} n^b$ transforms as 
\begin{align}
 \nn P^{\prime}{}^{bd} - n^{\prime}{}^b n^{\prime} {}^d = &
\frac{1}{u^2} \left ( \nn P^{bd} - n^b n^d \right), \label{Pnn1} \\
 P^{\prime}{}^{ab} n^{\prime}{} ^c - P^{\prime}{}^{ac} n^{\prime}{}^b = &
\frac{1}{u} \left (P^{ab} n^c - P^{ac} n^b 
+ \left ( \nn P^{ac} - n^a n^c \right )  V^b
- \left ( \nn P^{ab} - n^a n^b \right )  V^c
\right ).
\label{Pnn2}
\end{align}
\end{lemma}

\vs 

\noindent {\it Proof.} Let $Q^{ab} = 
\nn P^{ab} - n^a n^b$. According to Proposition \ref{TransContractions}, 
we start calculating 
$H_{c}^{\,\,b} \defi Q^{ab} \gamma_{ac} = 
\left ( \nn P^{ab} - n^a n^b \right ) \gamma_{ac} =
\nn \delta^b_{c}$ and $R^b \defi Q^{ab} \ell_a = - n^b$. Consequently
\begin{align*}
& \hat{H}_{c}^{\,\,\,b} = \nnprime \delta^b_c = \frac{\nn}{u^2} \delta^b_c, \\
& \hat{R}^b = - n^{\prime}{}^{b} =  - \frac{1}{u} \left ( n^b - V^b \nn
\right ).
\end{align*}
It follows from (\ref{id1})-(\ref{id2}) that $\hat{Q}^{ab}$ satisfies
\begin{align*}
&\hat{Q}^{ab} \gamma_{ac} = \hat{H}_c^{\,\,b} = \frac{\nn}{u^2} \delta^b_{c}, \\
& \hat{Q}^{ab} \ell_{a} = \frac{\hat{R}^b}{u} - V^c \hat{H}_c^{\,\,b} =
- \frac{1}{u^2} n^a + \frac{1}{u^2} V^b \nn - V^c \frac{\nn}{u^2} \delta^b_c
= - \frac{1}{u^2} n^a.
\end{align*}
Comparing with the expressions for $H_a^{\,\,b}$ and $R^b$ above (or,
alternatively, using Lemma \ref{raising}), it follows
immediately that $\hat{Q}^{ab} = \frac{1}{u^2} \left ( \nn P^{ab}
- n^a n^b \right )$. This establishes (\ref{Pnn1}).
For the second statement, let
\begin{eqnarray*}
H_{d}^{\,\,bc} \defi \left ( P^{ab} n^c  - P^{ac} n^b \right ) \gamma_{ad} =
%(  \delta^b_d n^c - n^b \ell_d n^c - \delta^c_d n^b + n^c \ell_d n^b \right ) = 
(  \delta^b_d n^c - \delta^c_d n^b ), \quad  \quad
R^{bc} \defi 
\left ( P^{ab} n^c  - P^{ac} n^b \right ) \ell_{a} = 0.
\end{eqnarray*}
Equations (\ref{id1})  and (\ref{id2}) become
\begin{align*}
& \hat{H}_d^{\,\,bc} =  \delta^b_d n^{\prime} {}^c - \delta^c_d n^{\prime} {}^b 
= \frac{1}{u} \left ( 
\delta^b_d n^c - \nn \delta^b_d V^c - \delta^c_d n^b +
\nn \delta^c_b V^b \right )  
= \hat{Q}^{abc} \gamma_{ad}, \\
& \hat{R}^{bc} = 0 = u \left ( \hat{Q}^{abc} \ell_{a} + V^a \hat{H}_a^{\,\,bc}
\right ) = u \hat{Q}^{abc} \ell_a + V^b n^c - V^c n^b 
\end{align*}
Using now Lemma \ref{raising} gives
$Q^{abc} = \frac{1}{u} \left (P^{ab} n^c - P^{ac} n^b 
+ \left ( \nn P^{ac} - n^a n^c \right )  V^b
- \left ( \nn P^{ab} - n^a n^b \right )  V^c
\right )$, which proves (\ref{Pnn2}). \ep

\section{Einstein tensor on the hypersurface.}
\label{constraints}

In this section we obtain the constraint equations 
in the case of general hypersurfaces. 
%This also generalizes
%and clarifies an expression obtained in \cite{} for 
%$G^{\mu}_{\,\,\nu} n_{mu} e_c^{\nu}$ 
%in the case of a null hypersurface.
We start with the following lemma, which shows that the components
$G^{\mu}_{\,\,\nu} n_{\mu} \ell^{\nu}$ and
$G^{\mu}_{\,\,\nu} n_{\mu} e_c^{\nu}$  can be obtained
in terms of the hypersurface data whenever this data is embedded
in a spacetime (our sign conventions for the Riemann, Ricci
and Einstein tensor follow \cite{Wald}).

\begin{proposition}
\label{EinRiem}
Let $\{ \N,\gamma,\ellc,\ll,{\bm Y} \}$ be 
embedded hypersurface data with embedding
$\Phi$ and rigging $\ell$. Let $\{ \hat{e}_a \}$ be a basis of $T \N$ and $e_a \defi \Phi_{\star} (\hat{e}_a)$.
Then
\begin{align}
 G^{\mu}_{\,\,\nu} n_{\mu} \ell^{\nu} & \eqN 
- R_{\alpha\beta\gamma\delta} 
\ell^{\alpha} e_b^{\beta} e_{c}^{\gamma} e_d^{\delta} n^c P^{bd} - 
\frac{1}{2} 
R_{\alpha\beta\gamma\delta}  e_a^{\alpha} e_b^{\beta} e_{c}^{\gamma} e_d^{\delta} 
P^{ac} P^{bd}, \label{Gnl} \\
G^{\mu}_{\,\,\nu} n_{\mu} e_c^{\nu} & \eqN 
 R_{\alpha\beta\gamma\delta}
\ell^{\alpha} e_b^{\beta} e_{c}^{\gamma} e_d^{\delta} 
\left ( \nn P^{bd} - n^b n^d\right )
+ R_{\alpha\beta\gamma\delta}  e_a^{\alpha} e_b^{\beta} e_{c}^{\gamma} e_d^{\delta} n^a 
P^{bd}. \label{Gne}
\end{align}
where $R^{\alpha}_{\,\,\,\beta\gamma\delta}$ is the Riemann tensor
of the ambient spacetime $(\M,\gM)$
and  $G^{\mu}_{\,\,\,\nu}$ is the corresponding Einstein tensor.
\end{proposition}

\noindent {\it Proof.} We start by obtaining an expression for the spacetime metric $\gM^{\alpha\gamma}$ in terms
of the data on $\N$. Since $\{ e_a, \ell \}$ is a  basis of $T_{\N} \M$ and
$\{ \bmomega_a, \bm{n} \}$ is the dual basis we can decompose
$\gM^{\alpha\gamma}$ as (see (\ref{ginv}))
\begin{eqnarray*}
\gM^{\alpha\gamma} \eqN P^{ab} e_a^{\alpha} e_b^{\gamma} 
+ n^a \left ( e_a^{\alpha} \ell^{\gamma} + \ell^{\alpha} e_a^{\gamma} 
\right ) + \nn \ell^{\alpha} \ell^{\gamma}. 
\end{eqnarray*}
Using  $n^a e_a^{\alpha} = n^{\alpha} - \nn \ell^{\alpha}$ the following
alternative expression also holds
\begin{eqnarray}
\gM^{\alpha\gamma} \eqN P^{ac} e^{\alpha}_a e^{\gamma}_c +  \left ( 
n^\alpha \ell^\gamma+ n^{\gamma} \ell^{\alpha} \right ) - \nn 
\ell^{\alpha} \ell^{\gamma}. \label{metricup}
\end{eqnarray}
We now calculate the curvature scalar
$R = \gM^{\beta\delta} R_{\beta\delta} = \gM^{\alpha\gamma} \gM^{\beta \delta} R_{\alpha\beta
\gamma\delta}$.  Inserting (\ref{metricup}) and using the symmetries of the
Riemann tensor yields
\begin{eqnarray}
\label{curvature}
R \eqN R_{\alpha\beta\gamma\delta}  e_a^{\alpha} e_b^{\beta} e_{c}^{\gamma} e_d^{\delta} 
P^{ac} P^{bd} 
+ 4  R_{\alpha\beta\gamma\delta} \ell^{\alpha} e_b^{\beta} n^{\gamma} e_d^{\delta} P^{bd} 
- 2 \nn 
R_{\alpha\beta\gamma\delta} \ell^{\alpha} e_b^{\beta} \ell^{\gamma} e_d^{\delta} P^{bd} 
- 2 R_{\alpha\beta\gamma\delta} \ell^{\alpha} n^{\beta} \ell^{\gamma}
n^{\delta}.
\end{eqnarray}
We can now calculate $G^{\mu}_{\,\,\nu} n_{\mu} \ell^{\nu} \eqN
R_{\alpha\beta\gamma\delta} n^{\beta} \ell^{\delta} \gM^{\alpha\gamma} 
- \frac{1}{2} R$. Inserting (\ref{metricup}), the first term is
\begin{eqnarray}
\label{firstterm}
R_{\alpha\beta\gamma\delta} n^{\beta} \ell^{\delta} \gM^{\alpha\gamma}  \eqN
R_{\alpha\beta\gamma\delta} \ell^{\alpha} e_b^{\beta} n^{\gamma} e_d^{\delta} P^{bd} 
- R_{\alpha\beta\gamma\delta} \ell^{\alpha} n^{\beta} \ell^{\gamma}
n^{\delta}.
\end{eqnarray}
Subtracting (\ref{firstterm}) and $\frac{1}{2}  R$ (from (\ref{curvature})), it follows
\begin{eqnarray*}
G^{\mu}_{\,\,\nu} n_{\mu} \ell^{\nu} \eqN
- \frac{1}{2} R_{\alpha\beta\gamma\delta}  e_a^{\alpha} e_b^{\beta} 
e_{c}^{\gamma} e_d^{\delta}  P^{ac} P^{bd}
-  R_{\alpha\beta\gamma\delta} \ell^{\alpha} e_b^{\beta} n^{\gamma} e_d^{\delta} P^{bd} 
+ \nn
R_{\alpha\beta\gamma\delta} \ell^{\alpha} e_b^{\beta} \ell^{\gamma} e_d^{\delta} P^{bd} 
\end{eqnarray*}
Using here $n^{\gamma} = n^c e_c^{\gamma} + \nn \ell^{\gamma}$ yields (\ref{Gnl}).

Regarding (\ref{Gne}), we need to evaluate 
$G^\mu_{\,\,\nu} n_{\mu} e_c^{\nu} =
R^{\mu}_{\,\,\nu} n_{\mu} e_c^{\nu} \eqN R_{\alpha\beta\gamma\delta} n^{\alpha} e_c^{\gamma} \gM^{\beta\delta}$.
Using (\ref{metricup}) we get
\begin{eqnarray*}
G^\mu_{\,\,\nu} n_{\mu} e_c^{\nu} \eqN
R_{\alpha\beta\gamma\delta} n^{\alpha} e_b^{\beta} e_c^{\gamma} e_d^{\delta} P^{bd}
+ R_{\alpha\beta\gamma\delta}  n^{\alpha} \ell^{\beta} e_c^{\gamma}
\left ( n^{\delta} - \nn \ell^{\delta} \right ),
\end{eqnarray*}
which becomes  (\ref{Gne}) after inserting $n^{\alpha} = n^c e_c^{\alpha} + \nn \ell^{\alpha}$. \ep

\vs

\noindent {\bf  Remark \arabic{remark}. \addtocounter{remark}{1}}  $G^{\mu}_{\,\,\nu} n_{\mu} e_c^{\nu}$ could
also have been obtained by exploiting the 
gauge transformations of Section \ref{gauge}. Inserting the transformation (\ref{gaugel})
into (\ref{Gnl}) and using the fact that $V^a$ is arbitrary implies the validity of (\ref{Gne}). 
This method is longer but straightforward and provides a non-trivial consistency check both 
for the gauge transformations
and for the expressions (\ref{Gnl}) and (\ref{Gne}).

\vs

Our next aim is to write down the right-hand sides of (\ref{Gnl}) and (\ref{Gne}) in terms of
hypersurface data. We start with the following identity, which has been obtained in \cite{MarsSenovilla1993} (see formulas (12) and (13) 
there), and which can be proved by inserting the decomposition (\ref{covder1})
in the Ricci identity for the Riemann tensor of the ambient spacetime.

\begin{proposition}[Mars \& Senovilla \cite{MarsSenovilla1993}]
\label{GaussCodazzi}
Let $\{ \N,\gamma,\ellc,\ll,{\bm Y} \}$ be embedded hypersurface data with embedding
$\Phi$ and rigging $\ell$. Let $\{ \hat{e}_a \}$ be a basis of $T \N$ and $e_a \defi \Phi_{\star} (\hat{e}_a)$.
Then
\begin{eqnarray}
R^{\alpha}_{\,\,\,\beta\gamma\delta} e^{\beta}_b e^{\gamma}_c e^{\delta}_d \eqN 
\left ( \Riemb^{a}_{\,\,\,bcd} - \Kn_{bd} \Psi^{a}_{\,\,\,c} + \Kn_{bc} \Psi^{a}_{\,\,\,d} \right) e_a^{\alpha} 
- \left ( \nablab_{c} \Kn_{bd} - \nablab_{d} \Kn_{bc} + \Kn_{bd} \varphi_c - \Kn_{bc} \varphi_{d} \right ) \ell^{\alpha},
\label{Riemeee}
\end{eqnarray}
where $R^{\alpha}_{\,\,\,\beta\gamma\delta}$ is the Riemann tensor of the ambient spacetime $(\M,\gM)$
and $\Riemb^{a}_{\,\,\,bcd}$ the curvature tensor of $\Gamb^{a}_{ab}$ in the basis $\{ \hat{e}_a \}$.
\end{proposition}
This result has the following corollary, which in particular defines two quantities denoted by (I)
and (II), which will play a useful role later.
\begin{corollary}
\label{corollaryRleee}
With the same hypotheses as in Proposition \ref{GaussCodazzi}, the following identities  hold
\begin{align}
R_{\alpha\beta\gamma\delta} \ell^{\alpha} e^{\beta}_b e^{\gamma}_c e^{\delta}_d & \eqN 
\ell_a \Riemb^{a}_{\,\,\,bcd} + \ll \left ( \nablab_{d} \Kn_{bc} -\nablab_{c} \Kn_{bd} 
\right ) + \frac{1}{2} \Kn_{bc} \nablab_{d} \ll - 
\frac{1}{2} \Kn_{bd} \nablab_c \ll \defi (I). \label{Riemleee} \\
R_{\alpha\beta\gamma\delta} e^{\alpha}_a e^{\beta}_b e^{\gamma}_c e^{\delta}_d & \eqN
\gamma_{af} \Riemb^f_{\,\,\,bcd} - \Kn_{bd} \left (\nablab_c \ell_a + \ll \Kn_{ca} \right )
+ \Kn_{bc} \left ( \nablab_{d} \ell_a + \ll \Kn_{da} \right ) - \ell_a
\left ( \nablab_c \Kn_{bd} - \nablab_d \Kn_{bc} \right ) \defi (II).  
\label{Riemeeee}
\end{align}
\label{FundId}
\end{corollary}

\noindent {\it Proof.} For (\ref{Riemleee}), contract (\ref{Riemeee}) with $\ell_{\alpha}$ and use equation (\ref{comp3}). 
For (\ref{Riemeeee})
contract the same expression with $e^{\alpha}_a$ and use (\ref{comp2}). 
\ep

\vs 

\noindent {\bf Remark \arabic{remark}. \addtocounter{remark}{1}} Identity (\ref{Riemleee}) also 
appears, in a slightly modified
form, in expression (17) of \cite{MarsSenovilla1993}.

\vs

We are now in a position where the components 
$G^{\mu}_{\,\,\nu} n_{\mu} \ell^{\nu}$, 
$G^{\mu}_{\,\,\nu} n_{\mu} e_c^{\nu}$ of the Einstein tensor can be obtained in terms of hypersurface data

\begin{theorem} 
\label{teor1}
Let $\{ \N,\gamma,\ellc,\ll,{\bm Y} \}$ be embedded hypersurface data with embedding
$\Phi$, rigging $\ell$ and ambient spacetime $(\M,\gM)$. Denote by 
$G^{\mu}_{\,\,\nu}$ the Einstein tensor of $(\M,\gM)$. Let $\{ \hat{e}_a \}$ be a basis of $T \N$ and $e_a \defi \Phi_{\star} (\hat{e}_a)$.
Then the following identities hold
\begin{align}
2 G^{\mu}_{\,\,\nu} n_{\mu} \ell^{\nu}  \eqN &
- P^{bd} \Riemb^{c}_{\,\,\,bcd} 
- P^{bd} n^c \left ( 
\ell_a \Riemb^{a}_{\,\,\,bcd}+
\nablab_d \left ( \ll \Kn_{bc} \right ) -
 \nablab_c \left ( \ll \Kn_{bd} \right ) \right )  - P^{ac} P^{bd} \left ( \Kn_{bc} Y_{da} -  \Kn_{bd} Y_{ac}  \right )
\label{EinnlHypData} \\
G^{\mu}_{\,\,\nu} n_{\mu} e_c^{\nu} \eqN & 
\left (P^{bd} - \ll n^b n^d \right )
\left ( 
\nablab_d \Kn_{bc} 
- \nablab_c \Kn_{bd} 
\right )
+ \frac{1}{2} \left ( \nn P^{bd} - n^b n^d \right ) 
\left (  \Kn_{bc} \nablab_d \ll  - \Kn_{bd} \nablab_c \ll \right)+ \nonumber \\ 
& +  P^{bd} n^a \left [ \Kn_{bc} \left ( Y_{da} + F_{da}  \right ) 
- \Kn_{bd}  \left ( Y_{ca} + F_{ca} \right ) 
 \right ] - n^b n^d \ell_a \Riemb{}^a_{\,\,\,bcd}.
\label{EinneHypData} 
\end{align}
\end{theorem}

\vs

\noindent {\it Proof.} Proposition \ref{EinRiem} shows, in particular, that
$G^{\mu}_{\,\,\nu} n_{\mu} \ell^{\nu} = -P^{bd} X_{bd}$, where 
\begin{eqnarray*}
X_{bd} \defi  \left . \left ( R_{\alpha\beta\gamma\delta} 
\ell^{\alpha} e_b^{\beta} e_{c}^{\gamma} e_d^{\delta} n^c  +
\frac{1}{2} 
R_{\alpha\beta\gamma\delta}  e_a^{\alpha} e_b^{\beta} e_{c}^{\gamma} e_d^{\delta} 
P^{ac}  \right ) \right  |_{\N}.
\end{eqnarray*}
So, we start finding an expression for $X_{bd}$ in terms of hypersurface data. Using (\ref{EqP1})
and (\ref{EqP4}) in (\ref{Riemeeee}) yields 
immediately
\begin{eqnarray}
R_{\alpha\beta\gamma\delta}  e_a^{\alpha} e_b^{\beta} e_{c}^{\gamma} e_d^{\delta} 
P^{ac} = \Riemb^c_{\,\,\,bcd} - n^c \ell_a \Riemb^{a}_{\,\,\,bcd} - \Kn_{bd} P^{ac} Y_{ac} + \Kn_{bc} P^{ac}
\left ( Y_{da} + F_{da} \right ) + \ll n^c \left ( \nablab_c \Kn_{bd} - \nablab_d \Kn_{bc} \right )
\label{RiemeeeeP}
\end{eqnarray}
where we used $\nablab_c \ell_a + \ll \Kn_{ca} = Y_{ca} + F_{ca}$ (see (\ref{nablabell}))
and the fact that $F_{ca}$ is antisymmetric. Contracting (\ref{Riemleee}) with $n^c$ and 
combining with (\ref{RiemeeeeP}) into $X_{bd}$ implies easily
\begin{eqnarray*}
X_{bd} = \frac{1}{2} \left ( \frac{}{} \Riemb^{c}_{\,\,\,bcd} + n^c \ell_a \Riemb^{a}_{\,\,\,bcd}
+ P^{ac} \Kn_{bc} \left ( Y_{da} + F_{da} \right ) - P^{ac} \Kn_{bd} Y_{ac} + n^c 
\left [ \nabla_d \left ( \ll \Kn_{bc} \right ) -
 \nablab_c \left ( \ll \Kn_{bd} \right ) \right ] \right ).
\end{eqnarray*}
Contracting this with $P^{bd}$, (\ref{EinnlHypData}) follows directly after using $P^{ac} P^{bd} \Kn_{bc} F_{da} =0$,
which holds because it is the contraction
of a symmetric and an antisymmetric tensor.

To prove (\ref{EinneHypData}) we evaluate first
\begin{align}
Z_{bcd} & \defi R_{\alpha\beta\gamma\delta} e_a^{\alpha} e_b^{\beta} e_c^{\gamma} e_d^{\delta} n^a  + \nn
R_{\alpha\beta\gamma\delta} \ell^{\alpha} e_b^{\beta} e_c^{\gamma} e_d^{\delta}  
 = n^a \left [ \Kn_{bc} \left ( Y_{da} + F_{da} \right )
- \Kn_{bd}\left ( Y_{ca} + F_{ca} \right ) \right ]
+ \nonumber \\
& + \nabla_d \Kn_{bc} - \nablab_c \Kn_{bd} 
+ \frac{\nn}{2} \left ( \Kn_{bc} \nablab_d \ll - 
\Kn_{bd} \nablab_c \ll \right ). \label{Z}
\end{align}
where we used Corollary \ref{FundId} and (\ref{EqP4}), (\ref{EqP3}).
Identity 
(\ref{EinneHypData}) now follows from (see (\ref{Gne}))
\begin{eqnarray*}
G^{\mu}_{\,\,\,\nu} n_{\mu} e_c^{\nu} = 
P^{bd} Z_{bcd} - n^b n^d R_{\alpha\beta\gamma\delta} \ell^{\alpha} e_b^{\beta} e_c^{\gamma} e_d^{\delta} 
\end{eqnarray*}
after using (\ref{Riemleee}) and (\ref{Z}). \ep

\vs 
We have worked so far using the connection $\Gamb^{a}_{bc}$, which 
arises naturally from the embedding of the hypersurface data in a
spacetime. Note, however, that this connection depends on the full
hypersurface data, in particular on $Y_{ab}$. If one wishes to view
the constraint equations as field equations for the hypersurface data, this
mixture between the connection $\Gamb^a_{bc}$ and the variable $Y_{ab}$
obscures notably the equations. Note that this {\it does not} happen
in the usual constraint  equations for spacelike hypersurfaces, where
the connection on $\Sigma$ is fully independent of the extrinsic curvature.
It is natural to try and do something similar in the general context and
make all dependence on $Y_{ab}$ in the constraint
equations fully explicity.
We will find an example where this strategy is useful 
in Section \ref{ThinShells} below.

In order to accomplish this, it is necessary to introduce a connection
which does not depend on $Y_{ab}$. In view of (\ref{GambProp}), the natural choice
is the following.
\begin{definition}
\label{DefGamo}
Let $\{\N,\gamma,\ellc,\ll\}$ be hypersurface metric data. We define
the {\bf metric hypersurface connection} $\Gamo{}^{a}_{bc}$ by
\begin{eqnarray}
\Gamo{}^{c}_{ab} & \defi &
\frac{1}{2} P^{cd} \left ( \partial_a \gamma_{bd} 
+ \partial_b \gamma_{ad} - \partial_d \gamma_{ab} \right )
+ \frac{1}{2} n^c  \left (\partial_a \ell_b
+ \partial_b \ell_a \right ). 
\end{eqnarray}
\end{definition}

\vs

\noindent {\bf  Remark \arabic{remark}. \addtocounter{remark}{1}} Despite its name, the metric hypersurface connection is, in general,
not the Levi Civita connection of any metric on $\N$. Since  
hypersurface metric data does not define any canonical metric on $\N$,
the name should not be a source of confusion. 

\vs 

It is clear that the metric hypersurface connection is torsion-free and that it relates to
the induced connection $\Gamb^a_{bc}$ by
\begin{eqnarray}
\Gamb^{a}_{bc} = \Gamo{}^a_{bc} - n^a Y_{bc}. \label{changeconnection}
\end{eqnarray}
It is well-known that two connections ${}^{(1)}\Gamma^{a}_{bc}$ and ${}^{(2)}\Gamma^{a}_{bc}$ whose difference
tensor is  $C^{a}_{bc}  \defi {}^{(1)}\Gamma^{a}_{bc} - {}^{(2)}\Gamma^{a}_{bc}$ define respective curvature tensors
${}^{(1)}R^a_{\,\,bcd}$ and ${}^{(2)}R^a_{\,\,bcd}$ satisfying (see e.g. formula (7.5.8) in \cite{Wald})
${}^{(1)}R^{a}_{\,\,bcd} = {}^{(2)} R^{a}_{\,\,bcd} + {}^{(1)}\nabla_c C^a_{bd} - {}^{(1)} \nabla_d C^a_{bc}
+ C^a_{cf} C^f_{bd} - C^a_{df} C^f_{cb}$,
where ${}^{(1)}\nabla$ is the covariant derivative of ${}^{(1)}\Gamma^a_{bc}$.
Applying this to the metric hypersurface
and induced connections, we find
\begin{eqnarray}
\Riemb{}^f_{\,\,\,bcd} = \Riemo{}^f_{\,\,\,bcd} + \nablao_d \left ( n^f Y_{bc} \right ) - \nablao_c \left ( n^f Y_{bd} \right )
+ n^f n^a \left ( Y_{ca} Y_{bd} - Y_{da} Y_{bc} \right ).
\label{riems}
\end{eqnarray}
Our aim is to rewrite the expressions in Corollary \ref{corollaryRleee} in terms of the connection 
$\Gamo$ and its curvature tensor. For that we need to transform covariant derivatives with respect to
$\nablab$ into covariant derivatives with $\nablao$. We collect the necessary expression in the following Lemma
\begin{lemma}
\label{identitiesnablao}
With the definitions above, let, in addition, $Z_{ab}$ be an arbitrary tensor. Then the following identities hold
\begin{align}
\nablao_a \ell_b & = F_{ab} - \ll \U_{ab}, \label{nablaoll}\\
\nablao_{a} \gamma_{bc} & = - \ell_b \U_{ac} - \ell_c \U_{ab}, \label{nablaogamma} \\
\nablao_{a} P^{bc} & = - \left ( n^b P^{cf} + n^c P^{bf} \right ) F_{af} - n^b n^c \partial_a \ll, \label{nablaoP} \\
\nablao_a n^b & = - \nn n^b \partial_a \ll - \left( \nn P^{bf} + n^b n^f \right ) F_{af} + P^{bf} \U_{af},
\label{nablaon} \\
\ell_a \Riemb^a_{\,\,\,bcd} & = \ell_a \Riemo{}^a_{bcd} + \nablao_d \left [ \nl Y_{bc} \right ]
- \nablao_c \left [ \nl Y_{bd} \right ] +  \nonumber \\
& + n^f \left [ Y_{bd} \left ( \nl Y_{cf} + F_{cf} - \ll \U_{cf} \right ) 
- Y_{bc} \left ( \left ( n^a \ell_a \right ) Y_{df} + F_{df} - \ll \U_{df} \right ) \right ],
\label{Riemell} \\
\gamma_{af} \Riemb^f_{\,\,\,bcd} & = \gamma_{af} \Riemo{}^{f}_{\,\,\,bcd} 
+ \ell_a \left ( \nablao_c \left ( \nn Y_{bd} \right ) 
- \nablao_d \left ( \nn Y_{bc} \right ) \right )  
+ Y_{bc} \left ( \U_{da} - \nn F_{da} \right ) \nonumber \\
& - Y_{bd} \left ( \U_{ca} - \nn F_{ca} \right ) 
 + \ell_a n^{f} \left [ Y_{bc} \left ( \nn Y_{df} + \U_{df} \right ) - Y_{bd} \left (
\nn Y_{cf} + \U_{cf} \right ) \right ],  \label{Riemgamma} \\
\nablab_d Z_{bc} - \nablab_c Z_{bd} & = \nablao_d Z_{bc} - \nablao_c Z_{bd} + n^f \left (
Y_{bd} Z_{fc} - Y_{bc} Z_{fd} \right ), \label{nablaZ} 
\end{align}
where we have defined $\nl \defi  n^a \ell_a$ and
$\U_{ab} \defi  \frac{1}{2} \left ( \pounds_{\hat{n}} \gamma_{ab}  + \ell_a \partial_b \nn
+ \ell_b \partial_a \nn \right )$.
\end{lemma}

\noindent {\it Proof.} We start by noticing that the definition of $\U_{ab}$ allows to write (see (\ref{KnProp}))
\begin{eqnarray}
\Kn_{ab} = \nn Y_{ab} + \U_{ab}. \label{KU}
\end{eqnarray}
The relation $\Gamb^a_{ab} = \Gamo{}^a_{bc} - n^a Y_{bc}$ implies
\begin{align*}
\nablao_a \ell_b & = \nablab_{a} \ell_b - n^c Y_{ab} \ell_{c} = Y_{ab} + F_{ab} - \ll K_{ab} - \nl Y_{ab} =  F_{ab} - \ll \U_{ab},
\end{align*}
where in the second equality we have used identity (\ref{nablabell}) and in the third equality we used (\ref{KU}) and
(\ref{EqP3}). This proves (\ref{nablaoll}). 
In order to prove (\ref{nablaogamma}) we use again the transformation rule for covariant derivatives
\begin{align*}
\nablao_{a} \gamma_{bc} & = \nablab_a \gamma_{bc} - n^f Y_{ab} \gamma_{fc} - n^f Y_{ac} \gamma_{bf} = 
- \ell_b \Kn_{ac} - \ell_c \Kn_{ab} + \nn \ell_{c} Y_{ab} + \nn \ell_b Y_{ac}  =  - \ell_b \U_{ac} - \ell_c \U_{ab}.
\end{align*}
where in the third equality we used (\ref{comp1}) and (\ref{EqP4}) and in the last one 
we employed (\ref{KU}). Expression (\ref{nablaoP}) is proved similarly; first transform
the covariant derivative $\nablao$ into the covariant derivative $\nablab$ to  obtain
\begin{eqnarray*}
\nablao_{a} P^{bc} = \nablab_a P^{bc} + n^a Y_{af} P^{fc} + n^c Y_{af} P^{bf}.
\end{eqnarray*}
Inserting equation (\ref{eqsinv1}) for $\nablab_{a} P^{bc}$ and recalling the 
explicit expression for $\Psi^c_a$ in Proposition \ref{solving} yields the result.
Before proving (\ref{nablaon}) we rewrite the equation (\ref{eqsinv2})
for $\nablab_a n^b$ directly in
terms of hypersurface metric data. Substituting the expression for $\varphi_a$
and $\Psi^c_{\,\,\,a}$ given in Proposition \ref{solving}, as well as $\Kn_{ab} = \nn Y_{ab} + \U_{ab}$,
into (\ref{eqsinv2}) implies
\begin{eqnarray*}
\nablab_a n^b +   \nn n^b \partial_a \ll + n^b n^f \left ( Y_{af} + F_{af} \right )
+ P^{bf} \left ( \nn F_{af} - \U_{af} \right ) =0.
\end{eqnarray*}
Applying to this equation the transformation law of covariant derivatives induced by the change of
connection (\ref{changeconnection})  gives (\ref{nablaon}).
For (\ref{Riemell}), simply multiply (\ref{riems}) by $\ell_{f}$, 
use $\ell_f \nablao_d \left ( n^f Y_{bc} \right ) = 
\nablao_d \left ( \nl Y_{bc} \right ) - n^f Y_{bc} \nablao_d \ell_{f}$ (i.e. ``integrate by parts'')  and use (\ref{nablaoll}).
To address (\ref{Riemgamma}), contract (\ref{riems}) with $\gamma_{af}$
and ``integrate by parts'' $\gamma_{af}$ in the second and third terms. After
using equation (\ref{nablaogamma}) and (\ref{EqP4}) one obtains
\begin{eqnarray*}
\gamma_{af} \Riemb^f_{\,\,\,bcd}  = \gamma_{af} \Riemo{}^{f}_{\,\,\,bcd} 
+ \nablao_c \left ( \nn  \ell_a Y_{bd} \right ) 
- \nablao_d \left ( \nn \ell_a Y_{bc} \right ) 
+  \nl \left ( Y_{bc} \U_{da} - Y_{bd} \U_{ca} \right ) + \\
 + \ell_a n^{f} \left [ Y_{bc} \left ( \nn Y_{df} + \U_{df} \right ) - Y_{bd} \left (
\nn Y_{cf} + \U_{cf} \right ) \right ]. 
\end{eqnarray*}
Breaking the derivative terms $\nablao_{d} \left ( \nn \ell_a Y_{bc} \right )
= \nn Y_{bc} \nablao_{d} \ell_a + \ell_a \nablao_{d} 
\left ( \nn Y_{bc} \right )$ and using (\ref{nablabell}) yields the result.
Finally, identity (\ref{nablaZ}) follows directly from the transformation law
\begin{eqnarray*}
\nablab_b Z_{bc} = \nablao_{b} Z_{bc} + n^f \left ( Y_{bd} Z_{fc} +  Y_{cd} Z_{bf} \right )
\end{eqnarray*}
induced by the change of connection. \ep

\vs 

\noindent We can now rewrite Corollary \ref{corollaryRleee}
in terms of the connection metric hypersurface connection.
\begin{proposition}
\label{Riemnablao}
Let $\{ \N,\gamma,\ellc,\ll,{\bm Y} \}$ be embedded hypersurface data with embedding
$\Phi$ and rigging $\ell$. Let $\{ \hat{e}_a \}$ be a coordinate basis of $T \N$ and $e_a \defi \Phi_{\star} (\hat{e}_a)$.
Then
\begin{align*}
R_{\alpha\beta\gamma\delta} \ell^{\alpha} e^{\beta}_b e^{\gamma}_c e^{\delta}_d \eqN &  
\ell_a \Riemo{}^a_{\,\,\,bcd} + \nablao_d Y_{bc} 
- \nablao_c Y_{bd} + \ll \left (  \nablao_d U_{bc} 
- \nablao_c \U_{bd} \right ) +  \frac{1}{2} \left ( U_{bc} \partial_d \ll  - 
 U_{bd} \partial_c \ll \right ) + \\
& + Y_{bd} \left ( \left ( F_{cf} + Y_{cf} \right ) n^f + \frac{1}{2} \nn 
\partial_c \ll \right ) -
Y_{bc} \left ( \left ( F_{df} + Y_{df} \right ) n^f + \frac{1}{2} \nn 
\partial_d \ll \right ). \\
R_{\alpha\beta\gamma\delta} e^{\alpha}_a e^{\beta}_b e^{\gamma}_c e^{\delta}_d  \eqN &
\gamma_{af} \Riemo{}^f_{\,\,\,bcd}
+ \ell_a \left ( \nablao_d \U_{bc} - \nablao_{c} \U_{bd} \right )
+ Y_{bc} \U_{da} - Y_{bd} \U_{ca} + \U_{bc} Y_{da} - \U_{bd} Y_{ca} + \\
& + \nn \left ( Y_{bc} Y_{da} - Y_{bd} Y_{ca} \right ) + \U_{bc} F_{da}
- \U_{bd} F_{ca}.
\end{align*}
\end{proposition}

\vs 

\noindent {\bf  Remark \arabic{remark}. \addtocounter{remark}{1}} Although we have chosen to state this proposition in terms of embedded
hypersurface data, in fact the proof works directly at the hypersurface data level and
establishes the identities
\begin{eqnarray}
(I) = 
\ell_a \Riemo{}^a_{\,\,\,bcd} + \nablao_d Y_{bc} 
- \nablao_c Y_{bd} + \ll \left (  \nablao_d U_{bc} 
- \nablao_c \U_{bd} \right ) +  \frac{1}{2} \left ( U_{bc} \partial_d \ll  - 
 U_{bd} \partial_c \ll \right ) + \nonumber  \\
+ Y_{bd} \left ( \left ( F_{cf} + Y_{cf} \right ) n^f + \frac{1}{2} \nn 
\partial_c \ll \right ) -
Y_{bc} \left ( \left ( F_{df} + Y_{df} \right ) n^f + \frac{1}{2} \nn 
\partial_d \ll \right ). \label{left1} \\
(II) = 
\gamma_{af} \Riemo{}^f_{\,\,\,bcd}
+ \ell_a \left ( \nablao_d \U_{bc} - \nablao_{c} \U_{bd} \right )
+ Y_{bc} \U_{da} - Y_{bd} \U_{ca} + \U_{bc} Y_{da} - \U_{bd} Y_{ca} + \nonumber \\
 + \nn \left ( Y_{bc} Y_{da} - Y_{bd} Y_{ca} \right ) + \U_{bc} F_{da}
- \U_{bd} F_{ca}. \label{left2}
\end{eqnarray}
where (I) and  (II) were defined, respectively, as  the right-hand sides of 
(\ref{Riemleee}) and (\ref{Riemeeee}).
\vs

\noindent {\it Proof.} We start with the rewriting of (I) in terms of the
metric hypersurface connection. Introducing $\ll$ inside
the derivatives in the second and third terms yields
\begin{align*}
(I) = & \,\, 
\ell_a \Riemb^{a}_{\,\,\,bcd} + \nablab_{d} \left ( \ll \Kn_{bc} \right )  -\nablab_{c} \left ( \ll \Kn_{bd}
\right ) - \frac{1}{2} \Kn_{bc} \nablab_{d} \ll +  \frac{1}{2} \Kn_{bd} \nablab_c \ll = \\
= & \,\,  \ell_a \Riemo{}^a_{bcd} + \nablao_b \left [ \nl Y_{bc} + \ll \Kn_{bc} \right ]
- \nablao_c \left [ \nl Y_{bd} + \ll \Kn_{bf} \right ] 
+ \nonumber \\ 
& + n^f \left ( Y_{bd} Y_{cf} - Y_{bc} Y_{df} \right )
\left [ \nl + \nn \ll \right ] 
  n^f \left ( Y_{bd} F_{cf} - Y_{bc} F_{df} \right ) 
- \frac{1}{2} \Kn_{bc} \nablao_{d} \ll +  \frac{1}{2} \Kn_{bd} \nablao_c \ll,
\end{align*}
where in the second equality we have used (\ref{Riemell}) and (\ref{nablaZ}) applied
to $Z = \ll K$. Using now $\nl  + \nn \ll = 1$  and 
\begin{eqnarray*}
\nl Y_{bd} + \ll \Kn_{bd} = \nl Y_{bd} + \ll \nn Y_{bd} + \ll \U_{bd} = Y_{bd} + \ll \U_{bc}
\end{eqnarray*}
proves (\ref{left1}). Identity (\ref{left2}) is proved by direct
substitution of (\ref{Riemgamma})  
in the left-hand side of (\ref{Riemeeee}) and using 
$\nabla_c \ell_a + \ll K_{ca} = Y_{ca} + F_{ca} $
and applying identity (\ref{nablaZ}) to $K_{bc} = \nn Y_{bc} + \U_{bc}$. 
\ep

\vs 
In order to rewrite the constraint equations in terms of the metric hypersurface
connection the following expression are also required. 

\begin{lemma}
\label{furtheridentities}
With the same definitions as in Lemma \ref{identitiesnablao}, let $Z_{bc}$
by an arbitrary symmetric tensor. Then the following
expressions hold
\begin{align}
& \nablao_c \left ( \nn P^{bd} - n^b n^d \right )  = 
- \left ( \nn P^{bd} - n^b n^d \right ) \left ( 2 F_{cf} n^f  + \nn \partial_c \ll
\right ) + U_{cf} \left ( 2 n^f P^{bd} - P^{bf} n^d - P^{df} n^b
\right ), \label{derP-nn} \\
%\nablao_b \left ( \nn P^{bd} - n^b n^d \right )  =
%- 2 \nn P^{bd} F_{df} n^f -  \left ( \nn P^{bd} - n^b n^d \right )
%\nn \partial_c \ll
%+ 2 U_{df} \left ( n^f P^{bd} - P^{bf} n^d \right ), \\
& \nablao_d \left ( P^{bd} n^c - P^{bc} n^d \right )  = 
\nn \partial_d \ll \left ( P^{bc} n^d- P^{bd} n^c \right ) 
+ F_{df} \left ( n^b n^d P^{cf} - P^{bd} n^c n^f - \nn P^{bd} P^{cf} \right ) 
+ \nonumber \\
& \hspace{3mm}  + U_{df} \left ( P^{bd} P^{cf} - P^{bc} P^{df} \right ), 
\label{derPn}  \\
& \left ( \nablao_d Z_{bc} - \nablao_c Z_{bd} \right ) 
\left ( \nn P^{bd} - n^b n^d \right )
 =
\nablao_f \left [ \left ( \nn P^{bd} - n^b n^d \right ) \left (
\delta^f_{d} Z_{bc}  - \delta^f_c Z_{bd} \right ) \right ]
+ \left ( \nn P^{bd} - n^b n^d \right ) \times \nonumber \\
&  \hspace{3mm} \times \left [ Z_{bc} \left ( 2 F_{df} n^f + \nn \partial_d \ll \right )
- Z_{bd} \left ( 2 F_{cf} n^f + \nn \partial_c \ll \right ) \right ]
+ \left ( P^{df} n^b- P^{bd} n^f \right ) \left (
Z_{bc} U_{df} - 2 Z_{bd} U_{cf} \right ), \label{derZP-nn} \\
& \left ( \nablao_d Z_{bc} - \nablao_c Z_{bd} \right ) P^{bd} n^c
=  \nablao_d \left ( \left ( P^{bd} n^c - P^{bc} n^d \right ) Z_{bc} \right )
+ Z_{bc} \left [ \frac{}{}  U_{df} \left ( P^{bc} P^{df} - P^{bd} P^{cf} \right )
+ 2 F_{df} P^{bd} n^c n^f \right . \nonumber \\
& \hspace{3mm}  \left . \frac{}{}
 + \nn \partial_d \ll \left ( P^{bd} n^c- P^{bc} n^d \right )
\right ]. \label{derZPn}
\end{align}

\end{lemma}

\vs 

\noindent {\it Proof.} The first identity follows by expanding the products 
and using (\ref{nablaoP}) and (\ref{nablaon}). The term $\partial_c \nn$
is dealt with using the identity
\begin{eqnarray*}
\partial_c \nn = 2 U_{cf} n^f - 2 \nn F_{cf} n^f - (\nn){}^2 \partial_c \ll,
\end{eqnarray*}
which follows immediately from $K_{ab} = \nn Y_{ab} + U_{ab}$ and 
Corollary \ref{solving}. Identity (\ref{derPn}) is obtained after
a straightforward calculation using equations (\ref{nablaoP}) and (\ref{nablaon}).
For the third expression in the lemma, write
\begin{align*}
\left ( \nablao_d Z_{bc} - \nablao_c Z_{bd} \right ) & \left ( \nn P^{bd}
- n^b n^d \right )  = 
\left ( \nablao_f Z_{bh} \right ) \left ( \delta^f_{d} \delta^h_c - \delta^f_c \delta^h_d
\right ) \left ( \nn P^{bd} - n^b n^d \right )
= \\
& = \nablao_f \left [ \left ( \nn P^{bd} - n^b n^d \right )
\left ( \delta^f_d Z_{bc} - \delta^f_c Z_{bd} \right ) \right ]
- \left ( Z_{bc} \delta^f_d - Z_{bd} \delta^f_c \right )
\nablao_f \left ( \nn P^{bd} - n^b n^d \right ),
\end{align*}
where in the third equality we have ``integrated by parts'' the
factor $(\nn P^{bd} - n^b n^d )$.  Inserting now (\ref{derP-nn})
in the right hand side implies (\ref{derZP-nn}) after simple algebraic
simplifications. For the fourth identity, a simple renaming of indices gives
\begin{eqnarray*}
\left ( \nablao_d Z_{bc} - \nablao_c Z_{bd} \right ) P^{bd} n^c 
= \left ( P^{bd} n^c - P^{bc} n^d \right ) \nablao_d  Z_{bc}.
\end{eqnarray*}
Integrating by parts the factor
$(P^{bd} n^c - P^{bc} n^d)$ and using (\ref{derPn}) yields the result.
\ep

\vs

We can finally rewrite the constraint equations in terms of the
metric hypersurface connection.

\begin{theorem}
\label{teor2}
Let $\{ \N,\gamma,\ellc,\ll,{\bm Y} \}$ be embedded hypersurface data with embedding
$\Phi$, rigging $\ell$ and ambient spacetime $(\M,\gM)$. Denote by 
$G^{\mu}_{\,\,\nu}$ the Einstein tensor of $(\M,\gM)$. Let $\{ \hat{e}_a \}$ be a basis of $T \N$ and $e_a \defi \Phi_{\star} (\hat{e}_a)$.
Then the following identities hold
\begin{align}
- G^{\mu}_{\,\,\nu} n_{\mu} \ell^{\nu} \eqN &
 \frac{1}{2} \Riemo{}^{c}_{\,\,\,bcd} P^{bd} + \frac{1}{2}
\ell_a \Riemo{}^{a}_{\,\,\,bcd} P^{bd} n^c+
\nablao_d \left ( \left ( P^{bd} n^c - P^{bc} n^d \right ) Y_{bc}  \right )
 + \frac{1}{2} \nn P^{bd} P^{ac} \left ( Y_{bc} Y_{da} - Y_{bd} Y_{ca} \right )  
 \nonumber  \\
& +\frac{1}{2} \left (P^{bd} n^c - P^{bc} n^d \right )
\left [ \ll \nablao_d U_{bc}+
   \left ( U_{bc} + \nn Y_{bc}  \right ) \partial_d \ll 
+ 2 Y_{bc} \left ( F_{df} - Y_{df} \right ) n^f
\right ],
\label{EinnlHypData3}  \\
G^{\mu}_{\,\,\nu} n_{\mu} e_c^{\nu} \eqN & - \ell_a \Riemo{}^a_{\,\,\,bcd} n^b n^d
+ \nablao_f \left [ \left ( \nn P^{bd} - n^b n^d \right ) \left (
\delta^f_{d} Y_{bc}  - \delta^f_c Y_{bd} \right ) \right ] + 
 \left ( P^{bd} - \ll n^b n^d \right ) \left ( \nablao_d U_{bc} -
\nablao_c U_{bd} \right ) \nonumber \\ 
& 
+ \left ( \nn P^{bd} - n^b n^d \right ) \left [
 \frac{1}{2} \left ( U_{bc} + \nn Y_{bc} \right ) \partial_d \ll - \frac{1}{2}
\left ( U_{bd} + \nn Y_{bd} \right ) \partial_c \ll 
+ \left ( Y_{bc} F_{df}  - Y_{bd} F_{cf} \right ) n^f \right ] \nonumber \\
& + \left ( P^{bd} n^f - P^{bf} n^d \right ) Y_{bd} U_{cf}
+ P^{bd} n^f \left ( U_{bc} F_{df} - U_{bd} F_{cf} \right ).
\label{EinneHypData3} 
\end{align}
\end{theorem}

\vs

\noindent {\bf  Remark \arabic{remark}. \addtocounter{remark}{1}} At first sight these expressions look much more
complicated  than the corresponding expressions in Theorem \ref{teor1}. However, here the
dependence on the extrinsic part of the data $Y_{ab}$ is completely
explicit, while in Theorem \ref{teor1} several terms (as for instance 
$K_{ab}$ or the connection  $\nablab$) depend implicitly on this tensor. 

\vs 

\noindent {\it Proof. } According to Proposition
\ref{EinRiem} we need to compute
$ R_{\alpha\beta\gamma\delta} 
l^{\alpha} e_b^{\beta} e_{c}^{\gamma} e_d^{\delta} n^c P^{bd}  
+ \frac{1}{2} 
R_{\alpha\beta\gamma\delta}  e_a^{\alpha} e_b^{\beta} e_{c}^{\gamma} e_d^{\delta} 
P^{ac} P^{bd}$. i.e. $(I) P^{bd} n^c + (II) P^{ac} P^{bd}$ in the notation above.
For the first term, we use (\ref{left1}) and apply the identity (\ref{derZPn}) 
with $Z_{bc} = Y_{bc}$. This implies, after a number of cancellations,
\begin{align}
(I) P^{bd} n^c  = & \,\,  \ell_a \Riemo{}^a_{\,\,\,bcd} P^{bd} n^c 
+ \nablao_d \left [ 
\left ( P^{bd} n^c - P^{bc} n^d \right ) Y_{bc} 
\right ]
+ \left ( P^{bd} n^c - P^{bc} n^d \right )
\left ( \frac{1}{2} \left ( U_{bc} + \nn Y_{bc} \right ) \partial_d \ll 
+ \right . \nonumber \\
& \left . + \frac{}{} \ll \nablao_d U_{bc} + Y_{bc} \left ( F_{df} - Y_{df} \right ) n^f \right )
+ P^{bd} P^{cf} \left ( Y_{bd} U_{cf} - Y_{bc} U_{df} \right ).
\label{Gln1}
\end{align}
For the second term, we simply use $\gamma_{af} P^{ac} = \delta^c_{f} - n^c \ell_f$
(\ref{EqP1}) and use the antisymmetry of $F_{ab}$ to obtain
\begin{align}
\frac{1}{2} (II) P^{ac} P^{bd} = & 
\frac{1}{2} \Riemo{}^{c}_{\,\,\,bcd} P^{bd} -
\frac{1}{2} \ell_a \Riemo{}^a_{\,\,\,bcd} P^{bd} n^c + P^{bd} P^{cf} \left ( Y_{bd} \left ( U_{df} + \frac{1}{2} \nn Y_{df} 
\right ) 
- Y_{bd} \left ( U_{cf} + \frac{1}{2} \nn Y_{cf} \right ) \right )
\nonumber \\
& - \frac{1}{2} \ll \nablao_d U_{bc} \left ( P^{bd} n^c - P^{bc} n^d \right ).
\label{Gln2}
\end{align}
Adding (\ref{Gln1}) and (\ref{Gln2}) gives (\ref{EinnlHypData3}).

From Proposition \ref{EinRiem}, in order to obtain the expression for
$G^{\mu}_{\,\,\nu} n_{\mu} e_c^{\nu}$ we need to compute
$(I) (\nn P^{bd} - n^b n^d ) + (II) n^a P^b$.  The first term
follows directly from (\ref{left1}) and reads
\begin{align}
(I) \left (\nn P^{bd} - n^b n^d \right )  = &
\left ( \nn P^{bd} - n^b n^d \right )
\left [ \ell_a \Riemo{}^{a}_{\,\,\,bcd} + \nablao_d Y_{bc} - \nablao_c Y_{bd}
+ \ll \left ( \nablao_d U_{bc} - \nablao_c U_{bd} \right )
+ \frac{}{} \right . \nonumber \\
& \left . + \frac{1}{2} 
\left ( U_{bc} - \nn Y_{bc}   \right ) \partial_d \ll
- \frac{1}{2} \left ( U_{bd} - \nn Y_{bc}  \right ) \partial_c \ll
\right ] - n^b n^d Y_{bd} F_{cf} n^f \nonumber \\
& + \nn P^{bd} n^f \left [
Y_{bd} \left ( Y_{cf} + F_{cf} \right ) -
Y_{bc} \left ( Y_{df} + F_{df} \right ) \right].
\end{align}  
On the other hand (\ref{left2}) together with $n^a \gamma_{af} = -\nn \ell_{f}$
implies
\begin{align}
(II) \,  n^a P^{bd} = & - \nn \ell_a \Riemo{}^{a}_{\,\,\,bcd} P^{bd} 
+ \nl P^{bd} \left ( \nablao_d U_{bc} - \nablao_c U_{bd} \right )
+ P^{bd} n^f \left [
Y_{bc} \left ( U_{df} + \nn Y_{df} \right ) \right .  \nonumber \\
& \left .  - Y_{bd} \left ( U_{cf} 
+ \nn Y_{cf} \right ) + U_{bc} \left ( Y_{df} + F_{df} \right )
- U_{bd} \left ( Y_{cf} + U_{cf} \right ) \right ].
\end{align}
Adding the two and using $\nl + \ll \nn = 1$ implies
\begin{align}
G^{\mu}_{\,\,\nu} n_{\mu} e_c^{\nu}  \eqN & 
- \ell_a \Riemo{}^{a}_{bcd} n^b n^d
+ \left ( \nn P^{bd} - n^b n^d \right )
\left ( \nablao_d Y_{bc} - \nablao_c Y_{bd} \right )
+\left ( P^{bd} - \ll n^b n^d \right )
\left ( \nablao_d U_{bc} - \nablao_c U_{bd} \right )
+ \nonumber \\
& + \frac{1}{2} 
\left ( \nn P^{bd} - n^b n^d \right )
\left [ \left (U_{bc} - \nn Y_{bc} \right ) \partial_d \ll
- \left ( U_{bd} - \nn Y_{bd} \right ) \partial_c \ll \right ]
- n^b n^d Y_{bd} F_{cf} n^f \nonumber \\
& + P^{bd} n^f \left [ U_{bc} \left ( Y_{df} + F_{df} \right )
- U_{bd} \left ( Y_{cf} + F_{cf} \right )
+ Y_{bc} \left ( U_{df} - \nn F_{df}
\right ) - Y_{bd} \left ( U_{cf} - \nn F_{cf} \right ) \right ]. 
 \end{align}
We use now identity (\ref{derZP-nn}) with $Z_{bc} = Y_{bc}$ and simplify the 
resulting expression to obtain (\ref{EinneHypData3}). \ep

\vs

Having obtained the constraint equations for embedded hypersurface
data, we can now promote these identities to fields equations on the hypersurface data. To that aim, we introduce (still in the
embedded hypersurface case) a scalar $\rho_{\ell}$ and a one-form $J_a$
on $\N$ by the definitions 
\begin{eqnarray*}
\rho_{\ell} \defi - G^{\mu}_{\,\,\nu}  n_{\mu} \ell^{\nu},
\quad \quad
J_a \defi - G^{\mu}_{\,\,\nu}  n_{\mu} e_a^{\nu}.
\end{eqnarray*}
The sign in the definition of $\rho_{\ell}$ has been chosen
so that in the spacelike case and with
the standard choice of rigging $\ell^{\mu} = - n^{\mu}$ (recall
that $\ell^{\mu} n_{\mu} = 1$ throughout this paper) $\rho_{\ell}$
coincides with the energy-density measured by the observer
orthogonal to the hypersurface. By this analogy and the fact
that $\rho_{\ell}$ in any case measures the normal-transversal
component of the Einstein tensor, we call $\rho_{\ell}$ the
``energy along $\ell$''. Note, however, this  name is not intented to imply
any relation with a  physical energy measured by any spacetime
observer. Regarding $J_{a}$, this measures the energy flux in the case when
$n$ is timelike. Again we will refer to $J_a$ as ``energy
flux'' although in the general case $J_{a}$ does not correspond
to physical fluxes measured by any spacetime observer. Note that 
both $\rho_{\ell}$ and $J_a$ depend  on the choice of rigging. If we choose 
another rigging $\ell^{\prime} = u (\ell + V^a e_a)$ (which implies the
change $\bm{n}^{\prime} = \frac{1}{u} \bm{n}$), this objects 
transform as 
\begin{eqnarray*}
\rho_{\ell^{\prime}} = \rho_{\ell} + V^a J_a, \quad \quad
J^{\prime}_a = \frac{1}{u} J_a
\end{eqnarray*}
We put forward the definition of matter-hypersurface
data.

\begin{definition}[Matter-Hypersurface data]
An n-tuple
$\{\N,\gamma_{ab},\ell_{a},\ll,Y_{ab}, \rho_{\ell}, J_{a}\}$ formed by
 hypersurface data $\{\N,\gamma_{ab}$, $\ell_{a}$, $\ll, Y_{ab}  \}$,
a scalar  $\rho_{\ell}$ and a one-form $J_{a}$ on $\N$
is called {\bf matter-hypersurface data} provided $\rho_{\ell}$
and $J_a$ transform as
\begin{eqnarray*}
\rho_{\ell^{\prime}} = \rho_{\ell} + V^a J_a, \quad \quad
J^{\prime}_a = \frac{1}{u} J_a.
\end{eqnarray*}
under a gauge transformation defined by $(u, V^a)$ and the
following constraint field equations hold.
\begin{align}
\rho_{\ell} & =
 \frac{1}{2} \Riemo{}^{c}_{\,\,\,bcd} P^{bd} + \frac{1}{2}
\ell_a \Riemo{}^{a}_{\,\,\,bcd} P^{bd} n^c+
\nablao_d \left ( \left ( P^{bd} n^c - P^{bc} n^d \right ) Y_{bc}  \right )
 + \frac{1}{2} \nn P^{bd} P^{ac} \left ( Y_{bc} Y_{da} - Y_{bd} Y_{ca} \right )  
 \nonumber  \\
& +\frac{1}{2} \left (P^{bd} n^c - P^{bc} n^d \right )
\left [ \ll \nablao_d U_{bc}+
   \left ( U_{bc} + \nn Y_{bc}  \right ) \partial_d \ll 
+ 2 Y_{bc} \left ( F_{df} - Y_{df} \right ) n^f
\right ],
\label{EinnlHypData_a_3}  \\
J_a  & = 
   \ell_a \Riemo{}^a_{\,\,\,bcd} n^b n^d
- \nablao_f \left [ \left ( \nn P^{bd} - n^b n^d \right ) \left (
\delta^f_{d} Y_{bc}  - \delta^f_c Y_{bd} \right ) \right ] - 
 \left ( P^{bd} - \ll n^b n^d \right ) \left ( \nablao_d U_{bc} -
\nablao_c U_{bd} \right ) \nonumber \\ 
& 
- \left ( \nn P^{bd} - n^b n^d \right ) \left [
 \frac{1}{2} \left ( U_{bc} + \nn Y_{bc} \right ) \partial_d \ll - \frac{1}{2}
\left ( U_{bd} + \nn Y_{bd} \right ) \partial_c \ll 
+ \left ( Y_{bc} F_{df}  - Y_{bd} F_{cf} \right ) n^f \right ] \nonumber \\
& - \left ( P^{bd} n^f - P^{bf} n^d \right ) Y_{bd} U_{cf}
- P^{bd} n^f \left ( U_{bc} F_{df} - U_{bd} F_{cf} \right ).
\label{EinneHypData_b_3} 
\end{align}
\end{definition}

\vs 
For completeness, we also add
the definition of embedded matter-hypersurface data

\begin{definition}[Embedding of matter-hypersurface data]
Let $\{ \N,\gamma,\ellc,\ll,{\bm Y}, \rho_{\ell}, J_a \}$ be
matter-hy\-per\-sur\-face data. This data will be 
{\bf embedded} in a spacetime $(\M,\gM)$ if 
$\{ \N,\gamma,\ellc,\ll,{\bm Y}\}$ is embedded
with embedding $\Phi$ and rigging $\ell$ and, moreover
\begin{eqnarray*}
\rho_{\ell} = \Phi^{\star} \left ( G ( \ell, \bm{n}) \right ), \quad
\quad  \bm{J} = \Phi^{\star} ( G( \cdot, {\bm{n}}) ),
\end{eqnarray*}
where $G$ is the 1-covariant, 1-contravariant
Einstein tensor of $(\M,\gM)$ and $\bm{n}$ 
the one-form normal to $\N$ satisfying $\bm{n} (\ell)=1$.
\end{definition}

\section{Evolution equations for discontinuities of hypersurface
data.}
%\section{An application: Barrab\`es-Israel equations for arbitrary shells}
\label{ThinShells}

As discussed in the Introduction, the matching theory of two
spacetimes is a useful arena to apply the results above on
the constraint equations for general hypersurfaces. The ingredients
for the matching are two spacetimes 
$(\M^{\pm},\gMm^{\pm})$ with diffeomorphic boundaries
$\partial \M^{\pm}$. Let $\N$ be an abstract copy of $\partial \M^{+}$
(or $\partial \M^{-}$) and $\Phi^{\pm} : \N \rightarrow \M^{\pm}$ be embeddings 
such that $\Phi^{\pm} (\N) = \partial \M^{\pm}$. The first requirement for
a successful matching is that  the manifold constructed from the union of 
$\M^+$ and $\M^{-}$ with their boundaries identified via $\Phi^{+} \circ
(\Phi^{-}){}^{-1}$ admits an atlas and a continuous metric $g$ 
such that, when restricted to $\stackrel{\circ}{\M}{}^{\pm}$ (the interior of
each manifold with boundary) gives $\gMm^{\pm}$. The necessary and 
sufficient conditions for this to happen were studied first
by  Clarke and Dray \cite{ClarkeDray1987}
in the case of boundaries with constant
signature case (including null). Their result was that the necessary and 
sufficient condition was that the induced first fundamental form
on $\N$ from both embeddings $\Phi^{\pm}$ coincide.
It was noticed in 
\cite{MarsSenovilla1993} that the arguments extend without difficulty
to the case of arbitrary causal character. However, in the case when the
boundary has null points, the statement in Clarke and Dray is incomplete 
because the equality of the first fundamental form is a necessary condition
but it fails in general to be also sufficient. This was noticed in 
\cite{MarsSenovillaVera2007} where the necessary and sufficient conditions
were found. With the notation introduced above we can state
this theorem as follows
\begin{theorem}[\cite{ClarkeDray1987}, \cite{MarsSenovilla1993}, \cite{MarsSenovillaVera2007}]
\label{Match}
Consider two $(m+1)$-dimensional spacetimes $(\M^{\pm},g^{\pm})$ with
boundaries $\partial \M^{\pm} $. They 
can be matched across their boundaries to
produce a spacetime $(\M,g)$ with continuous metric (in a suitable differentiable
atlas) if and only if:
\begin{itemize}
\item[(i)] There exists hypersurface metric data $(\N,\gamma_{ab},\ell_a,\ll)$
which can be embedded both in $(\M^+,\gMm^+)$ and in
$(\M^-,\gMm^-)$ with respective embedding and riggings
$\Phi^{\pm}$ and $\ell^{\pm}$. Moreover, the embeddings satisfy
$\Phi^{\pm}(\N) = \partial \M^{\pm}$.
\item[(ii)] The rigging vectors $\ell^{\pm}$ point, respectively, inside 
and outside of $\M^{\pm}$.
\end{itemize}
\end{theorem}
It is clear that when the boundaries contain no null points, the 
equality of the induced metric is equivalent to items (i) and (ii)
(simply chose the rigging vector to be the unit normal pointing outwards
of the boundary in one spacetime and inwards in the other). When the
boundary admits null points (in particular, if it is null everywhere),
the existence of the rigging satisfying (i) and (ii) does not
follow \cite{MarsSenovillaVera2007}
from the equality of the first fundamental forms, so it needs to be
added to the statement of the theorem.

At the hypersurface data level, the requirements of this Theorem 
translate into the condition that 
we deal with two hypersurface data 
$(\N,\gamma_{ab}, \ell_a, \ll, Y^{\pm}_{ab})$ which differ at most
on the transverse tensor $Y^{\pm}_{ab}$, i.e. such that they define 
the same hypersurface metric data.

The next step in the matching procedure  is to analyze whether the 
spacetimes match without introducing any physical energy-momentum
tensor with support on the matching hypersurface. This was studied in
the spacelike case by  Darmois \cite{Darmois1927},
Lichnerowicz \cite{Lichnerowicz1955} and
O'Brien-Synge \cite{Synge1952}. Their proposals were different but a close relationship between them could be
established \cite{BonnorVickers1981}. The Darmois matching conditions are coordinate independent 
and  demand the coincidence of the second fundamental forms on each boundary. 
Given the relationship  (\ref{KnProp})
between $\Kn_{ab}$ and $Y_{ab}$ and using the fact that $\nn \neq 0$
in the nowhere null case, the Darmois matching conditions are equivalent to $Y^{+}_{ab} = 
Y^{-}_{ab}$ (c.f. \cite{MarsSenovilla1993}). When $[K_{ab} ] \equiv K^{+}_{ab} - K^{-}_{ab} =0$,
it follows that there is a subatlas of the matched spacetime  $(\M,g)$ where the metric $g$ is $C^1$
\cite{Israel1966, BonnorVickers1981}. 
The Riemann tensor of $(\M,\gMm)$ may be discontinuous
at $\N$ but it is otherwise regular everywhere. This is why, physically,
one concludes that there is no matter-energy, or gravitational field concentrated
on the matching hypersurface. On the other hand, if $[K_{ab}] \neq 0$ (still in the everywhere
spacelike case) then
the Riemann tensor viewed as a tensor distribution in $(\M,\gMm)$ ( see
\cite{Lichnerowicz1955, Taub1979, GerochTraschen1987, MarsSenovilla1993, LeFloch2007}
for details on how to define
and use tensor distributions in this setting)
has a Dirac delta function supported on $\N$. This is interpreted physically
as a layer of energy and momentum concentrated on the 
hypersurface $\N$ (a ``shell'' of matter-energy).
It turns out that 
the Dirac delta part of the Einstein tensor is (not yet leaving
 the nowhere null case) 
\begin{eqnarray*}
  {\cal G}^{\mu\nu} = \tau^{ab} e_a^{\mu} e_b^{\nu} \delta_{\cal N} \quad
\mbox{with} \quad 
\tau_{ab} = - \left ( [K_{ab} ] - [K] \gamma_{ab}
\right ), \quad \delta_{\cal N}: \mbox{ Dirac 
delta on } (\N,\gamma).
\end{eqnarray*}
where the Dirac delta distribution is defined by integration with the 
volume form of the induced metric on the shell.
The (distributional) conservation equations $\nabla_{\mu} G^{\mu}_{\,\,\,\nu}=0$ 
imply
\begin{eqnarray}
(K^{+}_{ab} + K^{-}_{ab} )  \tau^{ab} = 2  [ G_{\mu\nu} n^{\mu} n^{\mu} ],
\quad \quad
\nablab_b \tau^b_{\,\,a} =  [J_a],
\label{IsraelEqs}
\end{eqnarray}
which are the Israel field equations for the shell \cite{Lanczos1922, Lanczos1924, Israel1966}.
For the purposes of this paper, it is interesting
to note that these equations can be derived directly from the
constraint equations (\ref{density0})-(\ref{momentum0}) by simply taking the difference of both equations at
each side of the matching hypersurface, and using the fact that the induced metric and the corresponding Levi-Civita connection do not jump
across the shell. As mentioned in the Introduction, these field equations 
were extended to the null case by Barrab\`es and Israel 
\cite{BarrabesIsrael1991} with an argument based on taking
limits where the spacelike/timelike matching hypersurface becomes null. 

The matching theory
for hypersurfaces of arbitrary causal character was derived in \cite{MarsSenovilla1993}, 
where in particular an explicit expression was obtained
for the Einstein tensor
distribution of the matched spacetime $(\M,\gM)$ in the atlas where the metric $g$ is $C^0$.
The shell equations for matching hypersurfaces of arbitrary causal character can in principle be derived
from the distributional (contracted) Bianchi identities. However, having obtained
general expressions for the constraint equations of hypersurfaces of general causal character, we can also follow
a different strategy. Namely, we can obtain the shell equations by simply taking differences
of the constraint equations on two hypersurface data of the form
$(\N,\gamma_{ab}, \ell_a, \ll, Y^{\pm}_{ab})$. This is conceptually much
simpler, as there is no need to introduce spacetime distributions nor
specific atlas in the matched spacetime in order to perform the calculation. 
In addition,
it does not even  need to assume that a spacetime exists. This may seem spurious, but in fact it is not.
The reason is that initial data sets are, in principle, much more general than spacetimes  because
well-posedness is not to be expected in general if the data
has ``timelike'' points, in the sense that $\gamma_{ab}$ is of Lorentzian signature on some open subset.
Even when one expects a well-posedness theorem to hold (e.g. when $\gamma_{ab}$ is semipositive
definite), the actual result is still lacking and may well be a non-trivial task to prove
it. However, jump discontinuities 
on the data may still be considered and the field equations that they need to satisfy 
can be derived directly from the constraint equations. The important point is that
this makes all the sense already at the initial data level, and
may be useful for several things, ranging from studying shell equations on their own (i.e.
detached from the spacetime) to more practical purposes like obtaining new solutions
of the constraint equations from a seed solution of the constraints together 
with a solution of the shell equations on this metric hypersurface data background.

Let us therefore  try and find the shell equations from the results in Theorem \ref{constraints}.
Assume we are given two matter-hypersurface data 
$\{\N,\gamma_{ab},\ell_a,\ll,Y^{\pm}_{ab}, \rho^{\pm}_{\ell}, J^{\pm}_{a}\}$
and let us define $V_{ab} \defi Y^+_{ab} - Y^{-}_{ab}$. Let us also define
the jump in the ``energy density'' and in the ``energy flux'' as
$[\rho_{\ell}] \defi \rho^{+}_{\ell} - \rho^{-}_{\ell}$ and 
$[J_a] \defi J_a^{+} - J_a^{-}$.
When subtracting the constraint equations
for 
$\{\N,\gamma_{ab},\ell_a,\ll,Y^+_{ab}, \rho^+_{\ell}, J^+_{a}\}$
and for 
$\{\N,\gamma_{ab},\ell_a,\ll,Y^-_{ab}, \rho^-_{\ell}, J^-_{a}\}$
all terms that depend exclusively on the hypersurface metric data cancel
out. Furthermore, only one connection $\nablao$ appears in all equations,
since this depends only on the hypersurface metric data.
Hence, subtracting the constraint equations and using the trivial
identity
\begin{eqnarray*}
Y_{bc} Y_{da} - Y_{bd} Y_{ca} =
\Y_{bc} V_{da} + V_{bc} \Y_{da} - \Y_{bd} V_{ca} - V_{bd} \Y_{ca},
\end{eqnarray*}
where $\Y_{ab} \defi \onehalf  (Y^{+}_{ab} + Y^{-}_{ab} )$, gives
\begin{align}
[\rho_{\ell}] = & 
\nablao_f \left ( \left ( P^{bf} n^c - P^{bc} n^f \right ) 
V_{bc}  \right )
 + \frac{1}{2} \nn P^{bd} P^{ac} \left ( 
\Y_{bc} V_{da} + V_{bc} \Y_{da} - \Y_{bd} V_{ca} - V_{bd} \Y_{ca}
\right ) \nonumber  \\
& +\frac{1}{2} \left (P^{bd} n^c - P^{bc} n^d \right )
\left [ \nn V_{bc} \partial_d \ll 
+ 2 V_{bc} \left ( F_{df} - \Y_{df} \right ) n^f
- 2 \Y_{bc} V_{df} n^f  
\right ],
\label{EinnlHypData4}  \\
[J_{c} ] = & 
- \nablao_f \left [ \left ( \nn P^{bd} - n^b n^d \right ) \left (
\delta^f_{d} V_{bc}  - \delta^f_c V_{bd} \right ) \right ] + 
\nonumber \\ 
& 
- \left ( \nn P^{bd} - n^b n^d \right ) \left [
 \frac{1}{2} \nn V_{bc}  \partial_d \ll - \frac{1}{2} \nn
V_{bd}  \partial_c \ll 
+ \left ( V_{bc} F_{df}  - V_{bd} F_{cf} \right ) n^f \right ] \nonumber \\
& - \left ( P^{bd} n^f - P^{bf} n^d \right ) V_{bd} U_{cf}.
\label{EinneHypData4} 
\end{align}
Looking at the terms involving derivatives, we realize
that this expressions introduce naturally a 
$1-1$ tensor  $\tau^f_{\,\,c}$ and vector $T^d$  with the definitions
\begin{align*}
\tau^{f}_{\,\,c} \defi  & - \left ( \nn P^{bd} - n^b n^d \right ) \left (
 \delta^f_{d} V_{bc}  - \delta^f_c V_{bd} \right ), \\
T^f \defi  & \left ( P^{bf} n^c - P^{bc} n^f \right ) V_{bc}. 
\end{align*}
At the sight of Lemma \ref{raising} is it quite natural to enquire 
whether these tensors can be obtained, respectively, from contractions
of a suitable 2-contravariant tensor $\tau^{fa}$ with $\gamma_{ac}$ and
$\ell_a$. The result is given in the following Lemma.
\begin{lemma}
The tensors $\tau^{f}_{\,\,\,c}$ and $T^f$ defined above satisfy
$n^c \tau^{f}_{\,\,\,c} + \nn T^f \equiv 0$. Consequently, there
exists a tensor $\tau^{fa}$ such that $\tau^{fa} \gamma_{ac} =
\tau^{f}_{\,\,\,c} $ and $\tau^{fa} \ell_a =  T^a$. Moreover,
$\tau^{fa}$ takes the explicit form
\begin{eqnarray}
\tau^{fa} = \left ( n^f P^{ac} + n^a P^{fc} \right) n^d V_{cd} -
\left ( \nn P^{fc} P^{ad} + P^{fa} n^c n^d \right ) V_{cd}
 + \left ( \nn P^{fa} - n^f n^a \right ) P^{cd} V_{cd}.
\label{tau}
\end{eqnarray}
\end{lemma}

\vs

\noindent {\it Proof.} We compute
\begin{align*}
\tau^{f}_{\,\,\,c} n^c = &  
- \left ( \nn P^{bd} - n^b n^d \right )  \left ( \delta^f_d V_{bc} n^c -
n^f V_{bd} \right ) = - \nn P^{bf} V_{bc} n^c + n^b n^f V_{bc} n^c + 
\nn P^{bd} n^f V_{bd} - n^b n^d n^f V_{bd} = \\
= &  - \nn \left ( P^{bf}  n^c- P^{bc} n^f \right ) V_{bc} = - \nn T^f. 
\end{align*}
Lemma \ref{raising} applied to $Z_c \rightarrow \tau^{f}_{\,\,\,c}$
and $W \rightarrow T^f$implies immediately the existence of the tensor
$\tau^{fa}$ claimed in the statement. In order to
find its explicit expression we invoke again Lemma \ref{raising}, which gives
$\tau^{fa} = P^{ac} \tau^{f}_{\,\,\,c} + n^a T^f$. Expanding the 
right-hand side gives (\ref{tau}). \ep

The tensor $\tau^{af}$ has been defined by inspection of equations
(\ref{EinnlHypData4})-(\ref{EinneHypData4}). It turns out that this allows us to rewrite not only
the derivative terms of these equations in terms of $\tau^{fa}$ but
in fact the full equations as well. It is straightforward to
check that (\ref{EinnlHypData4})-(\ref{EinneHypData4}) can be rewritten in the form
\begin{align*}
  \nablao_{a} \left ( \tau^{ab} \ell_{b} \right )
+ \tau^{ab} \ell_{b} \left (  \frac{1}{2} \nn \partial_a \ll +  F_{ad} n^d  \right )
- \tau^{ab} \Y_{ab} = &   [ \rho_{\ell} ], \\
 \nablao_{b} \tau^{b}_{\,\,c} + \tau^{b}_{\,\,c}
  \left ( \frac{1}{2}  \nn \partial_b \ll + F_{bd} n^d \right ) 
+ \tau^{bd} \ell_d U_{bc} = &  [ J_c ].
\end{align*}
It is also immediate to check that, in the spacelike (or timelike case), these equations
become exactly the Israel equations (\ref{IsraelEqs}).

The tensor $\tau^{ab}$ has the following properties

\begin{proposition}
Let $\tau^{fa}$ be defined as in (\ref{tau}) and assume the hypersurface
data is of dimension $m \geq 2$. Then, the following
properties hold
\begin{itemize}
\item[(i)] $\tau^{fa}$ is symmetric, i.e. $\tau^{af} = \tau^{fa}$.
\item[(ii)] At any point $p \in \N$ where 
$\nn (p) \neq 0$, $\tau^{fa}$ vanishes if and only if $V_{ab} =0$, i.e.
if and only if the jump in $Y_{ab}$ vanishes. At 
any point where $\nn(p)=0$, 
$\tau^{fa}$ vanishes if an only if $V_{ab} n^b =0$ and
$P^{ab} V_{ab}=0$.
\item[(iii)] Under a gauge transformation of the data
$\{\N,\gamma_{ab},\ell_a,\ll,Y^{\pm}_{ab}\}$ defined by $\{ u, V^a \}$ we have 
$V'_{ab}  = u V_{ab}$ and 
\begin{eqnarray}
\label{gaugetau}
\tau^{\prime} {}^{fa} = \frac{1}{u} \tau^{fa}.
\end{eqnarray}
\item[(iv)] Consider ``non-degenerate data in the normal gauge'', i.e.
$\{ \N,\gamma_{ab},\ell_a =0,\ll = \epsilon, Y^{\pm}_{ab} = \epsilon
\Kn^{\pm}_{ab} \}$ with $\epsilon = \pm 1$ and $\gamma_{ab}$ of signature
$(+,\cdots, +, - \epsilon)$ . Then
\begin{eqnarray}
\tau^{ab} = - [ \Kn^{ab} ] + \gamma^{ab} [\Kn],
\label{nonnull}
\end{eqnarray}
where $[\Kn^{ab}] \defi \gamma^{ac} \gamma^{bd} \left ( \Kn^+_{cd} - \Kn^{-}_{cd}
\right )$, 
$[\Kn] \defi \gamma_{ab} [ \Kn^{ab} ]$
and $\gamma^{ab}$ is the inverse of $\gamma_{ab}$,
\end{itemize}
\end{proposition}

\vs

\noindent{\bf  Remark \arabic{remark}. \addtocounter{remark}{1}} The condition on the dimension is necessary
because in dimension $m=1$ (\ref{tau}) implies $\tau^{ab}
\equiv 0$.

 \vs

\noindent {\it Proof.} Item (i) is obvious from the explicit expression (\ref{tau}). For item (ii), we notice the $\tau^{fa}$ vanishes if and only
if $\tau^{f}_{\,\,\,c} =0$ and $T^f=0$. Taking the trace of the first
expressions gives
\begin{eqnarray}
0 = \tau^{f}_{\,\,f} = - (m-1) \left ( \nn P^{bd} - n^b n^d 
\right ) V_{bd}. \label{trace}
\end{eqnarray}
Lowering the index to $T^f$ implies
\begin{equation*}
0 = T^f \gamma_{fa} = V_{ac} n^c + \ell_a \left ( \nn P^{bc} - n^b n^c
\right ) V_{bc} = V_{ac} n^c,
\end{equation*}
where in the last equality we used (\ref{trace}). Inserting this back
into (\ref{trace}) gives $P^{bd} V_{bd} =0$. Using all this 
in $\tau^{f}_{\,\,a}$ implies
\begin{equation*}
0 = \tau^f_{\,\,\,c} = - \nn P^{bf} V_{bc}.
\end{equation*}
So far we have found that $\tau^{af}$ vanishes at one point if and
only if $P^{bd} V_{bd}=0$, $\nn P^{bf}V_{bc}=0$ and $V_{ac}n^c=0$. 
This establishes the result for points $p \in \N$ where $\nn(p)=0$.
For points where $\nn(p) \neq 0$, it only remains to show that
there exists no covector $S_a$ satisfying 
$P^{ab} S_{b} =0$ and $S_a n^a =0$ except for $S_a=0$. Any such covector 
satisfies
\begin{equation*}
\begin{bmatrix}
P^{ab} & n^a \\
n^b & \nn 
\end{bmatrix} 
\begin{bmatrix}
S_b  \\
0 
\end{bmatrix} 
= 
\begin{bmatrix}
0 \\
0 
\end{bmatrix}. 
\end{equation*}
Since the square matrix is invertible $S_b =0$ follows, as claimed.

For item (iii), the statement $V'_{ab} = u V_{ab}$ is immediate from the 
definition of $V_{ab} = Y^+_{ab} - Y^{-}_{ab}$ and the Definition
\ref{gauge} of gauge transformation. In order to address
the gauge behaviour of $\tau^{fa}$, Proposition
\ref{TransContractions} implies that 
(\ref{gaugetau}) is equivalent to 
\begin{align}
& \tau^{\prime}{}^f_{\,\,\,c} = \frac{1}{u} \tau^{f}_{\,\,\,c}, \nonumber \\
& T^{\prime}{}^f =  T^f + V^c \tau^f_{\,\,\,c}.
\label{break}
\end{align}
In order to prove these identities we need to compute the
transformation laws of $\tau^f_{\,\,\,c}$ and of $T^f$. 
The first one is a simple consequence of 
Lemma \ref{TranPnn} and reads
\begin{eqnarray*}
\tau^{\prime}{}^f_{\,\,\,c} = - \left ( 
\nnprime P^{\prime}{}^{bd} - n^{\prime} {}^b  n^{\prime} {}^d \right ) 
\left ( \delta^f_d V^{\prime}_{bc} - \delta^f_{c} V^{\prime}_{bd} \right ) = 
- \frac{1}{u^2} \left (
\nn P^{bd} - n^b  n^d \right ) 
u \left ( \delta^f_d V_{bc} - \delta^f_{c} V_{bd} \right ) = \frac{1}{u}
\tau^f_{\,\,\,c}.
\end{eqnarray*}
For the second, 
we use the second part of Lemma \ref{TranPnn} to obtain
\begin{align*}
T^{\prime}{}^f = &
\left (  P^{\prime}{}^{bf} n^{\prime}{}^c - P^{\prime}{}^{bc} n^{\prime}{}^f 
\right ) V_{bc}^{\prime} = 
 \left [P^{bf} n^c - P^{bc} n^f + \left ( \nn P^{bc} - n^b n^c
\right ) V^f - \left ( \nn P^{bf} - n^b n^f \right ) V^c 
\right ] V_{bc} = \\ 
= & T^f + V^a \left ( \nn P^{bc} - n^b n^c \right )
\left ( \delta^f_a V_{bc} - \delta^f_b V_{ac} \right )
= T^f + V^a \tau^f_a
\end{align*}
as desired. This proves claim (iii).

For claim (iv), first notice that $\ell_a =0$ and $\ll = \epsilon$
imply (from (\ref{EqP1})-(\ref{EqP4})) that $\nn = \epsilon$, $n^a=0$
and $P^{ab} = \gamma^{ab}$. Moreover, the expression 
$Y_{ab} = \epsilon \Kn_{ab}$ is consistent with (\ref{defKn}). Inserting
this information into the general expression for $\tau^{fa}$ yields
\begin{eqnarray*}
\tau^{fa} = \nn \left ( P^{fa} P^{cd} - P^{fc} P^{ad} \right ) V_{cd}
= \left ( \gamma^{fa} \gamma^{cd} - \gamma^{fc} \gamma^{ad} \right )
[\Kn_{cd}] = - [\Kn^{fa}] + \gamma^{fa} [K]. 
\end{eqnarray*}
\ep

\vs 

As a consequence of this proposition we find that the tensor $\tau^{ab}$ has the symmetries of
an energy-momentum tensor and coincides with the standard definition of the
energy-momentum tensor on the shell in the nowhere null case. Moreover, in the case
of matching across null
hypersurfaces, it is immediate to check that $\tau^{ab}$ coincides with the
definition of energy-momentum tensor put forward by Barrab\`es and Israel in expression (31)
in \cite{BarrabesIsrael1991}, where their tensor $g_{\star}^{ab}$ (defined by
$g_{\star}^{ab} \gamma_{bc} = \delta^a_c - n^a \ell_b$) is
related (in the null case)
to $P^{ab}$ by $g_{\star}^{ab} = P^{ab} + 2 \lambda n^a n^b$, for
an arbitrary $\lambda$. It can also be checked that,
in the case of matching across hypersurfaces of general causal character the tensor 
$\tau^{ab}e_a^{\mu} e_b^{\nu}$ agrees with the Dirac delta part of the Einstein tensor
of the matched spacetime $(\M,\gM)$ denoted by $\tau^{\alpha\beta}$ in \cite{MarsSenovilla1993} and given explicitly
in expression (71) in that reference. All these
considerations allow us to put forward the following definition, independently of whether the
hypersurface data is embedded in any spacetime or not.

\begin{definition}[Shell field equations]
A {\bf shell} is a pair of matter-hypersurface data
of the form
$\{\N$, $\gamma_{ab}$, $\ell_a$, $\ll$, $Y^{\pm}_a$, $\rho_{\ell}^{\pm}$,
$J^{\pm}_{c}\}$. 
The {\bf energy-momentum tensor} on the shell
is the  symmetric 2-covariant tensor $\tau^{ab}$ defined by
\begin{eqnarray*}
\tau^{ab} & \!\!\!\!\! \defi \!\!\! \!\!&
\left ( n^a P^{bc} + n^b P^{ac} \right) n^d V_{cd} -
\left ( \nn P^{ac} P^{bd} + P^{ab} n^c n^d \right ) V_{cd}
 + \left ( \nn P^{ab} - n^a n^b \right ) P^{cd} V_{cd},
\end{eqnarray*}
where $V_{ab} \defi Y^{+}_{ab} - Y^{-}_{ab}$. The {\bf shell field equations}
are the pair of partial differential equations
\begin{align}
  \nablao_{a} \left ( \tau^{ab} \ell_{b} \right )
+ \tau^{ab} \ell_{b} \left (  \frac{1}{2} \nn \partial_a \ll +  F_{ac} n^c  \right )
- \frac{1}{2} \tau^{ab} \left ( Y^+_{ab} + Y^-_{ab} \right ) = & [ \rho_{\ell} ], \label{shellrho} \\
 \nablao_{b} \tau^{b}_{\,\,a} + \tau^{b}_{\,\,a}
  \left ( \frac{1}{2}  \nn \partial_b \ll + F_{bc} n^c \right ) 
+ \tau^{bc} \ell_c U_{ba} =&  [ J_a ], \label{shellJ}
\end{align}
where $[\rho_{\ell}] \defi \rho^+_{\ell} - \rho^{-}_{\ell}$,
$[J_{a}] \defi J^{+}_a - J^{-}_a$  and
$\nablao$ is the metric hypersurface connection.
\end{definition}

\noindent {\bf Remark \arabic{remark}. \addtocounter{remark}{1}}
In the context of matching of spacetimes, it is costumary to use a connection
on $\Sigma$ which is the semi-sum of the corresponding rigging
connections from each side of the hypersurface. In other words,
it is costumary to use the connection
\begin{eqnarray*}
\hat{\Gamma}^a_{bc} \defi \frac{1}{2} \left ( \Gamb^{+}{}^a_{bc} + 
\Gamb^{-}{}^a_{bc} \right),
\end{eqnarray*}
where $\Gamb^{\pm}{}^{a}_{bc}$ is defined by expression (\ref{GambProp})
after substitution of $Y_{ab} \rightarrow Y^{\pm}_{ab}$. The relationship
between $\hat{\Gamma}^a_{bc}$ and $\Gamo{}^{a}_{ab}$ is,
obviously, $\Gamo{}^{a}_{ab} = \hat{\Gamma}^a_{bc} + n^a \overline{Y}_{bc}$.
A simple calculation allows us to rewrite (\ref{shellrho})-(\ref{shellJ}) in terms
of the covariant derivative $\hat{\nabla}$ associated to
$\hat{\Gamma}{}^a_{bc}$. The result (which  I add for easy of comparison between
our approach and the distributional approach in a matching context), is
\begin{align*}
\hat{\nabla}_a \left ( \tau^{ab} \ell_b \right )
+ \tau^{ab} \ell_b \left ( \frac{1}{2} \nn  \partial_a \ll + 
F_{ac} n^c + \overline{Y}_{ac} n^c \right )
- \tau^{ab} \overline{Y}_{ab} & = [ \rho_{\ell} ], \\
\hat{\nabla}_b \tau^b_{\,\,\,a} + \tau^{b}_{\,\,\,a} \left (
\frac{1}{2} \nn \partial_b \ll + F_{bc} n^c + \overline{Y}_{bc} n^c
\right ) + \tau^{bc} \ell_{c} \left ( \nn \overline{Y}_{ba} + U_{ba}
\right ) & = [ J_{a} ].
\end{align*}
\vs

We know that both the energy-momentum tensor $\tau^{ab}$ 
and $[J_{c}]$  have a very simple gauge behaviour. On the other hand,
the gauge behaviour of the connection
$\Gamo{}^c_{ab}$ is complicated. It makes sense to try and rewrite the equations above 
in such a form that the gauge dependence becomes explicit. In order to accomplish this,
it is convenient to work in coordinates. Indeed, the coordinate expression of (\ref{shellrho}) is
\begin{eqnarray*}
\partial_{a} T^a  + T^a \left (  \Gamo{}^c_{ac} +
\frac{1}{2} \nn \partial_a \ll +  F_{ac} n^c  \right )
- \frac{1}{2} \tau^{ab} \left ( Y^+_{ab} + Y^-_{ab} \right ) = & [ \rho_{\ell} ].
\end{eqnarray*}
so, we need to compute $\Gamo{}^c_{ac} + \frac{1}{2} \nn \partial_a \ll + F_{ac} n^c$. Recalling 
Definition \ref{DefGamo} we have
\begin{align*}
& \Gamo{}^c_{ac} = \frac{1}{2} P^{cd} \partial_a \gamma_{cd} + \frac{1}{2} n^c \left (
\partial_a \ell_c + \partial_c \ell_a \right ) \quad \Longrightarrow \quad \\
& \Gamo{}^{c}_{ac} + \frac{1}{2} \nn \partial_a \ll + F_{ac} n^c =
 \frac{1}{2} \left ( P^{cd} \partial_a \gamma_{cd} + n^c \partial_a \ell_c + \nn \partial_a \ll \right )
= \frac{1}{2} \mbox{tr} \left ( \mathbb{A}^{-1} \partial_a \mathbb{A} \right ),
\end{align*}
where the matrix $\mathbb{A}$ was introduced in Definition \ref{hypmetdata} (and we have used 
property (\ref{inverse})). We conclude therefore that
\begin{eqnarray*}
\Gamo{}^c_{ac}  + \frac{1}{2} \nn \partial_a \ll + F_{ac} n^c = \frac{1}{2 (\det \mathbb{A})}
\partial_a (\det \mathbb{A})
\end{eqnarray*}
and the field equation (\ref{shellrho}) becomes
\begin{eqnarray*}
\frac{1}{\sqrt{|\det (\mathbb{A})|}} \partial_a \left ( \sqrt{| \det (\mathbb{A})|} 
\tau^{ab} \ell_b \right ) - \frac{1}{2} \tau^{ab} \left ( Y^+_{ab} + Y^-_{ab} \right ) = [ \rho_{\ell} ].
\end{eqnarray*}
We can also compute the coordinate expression of equation (\ref{shellJ}):
\begin{align*}
 \partial_{b} \tau^{b}_{\,\,a} 
- \Gamo{}^c_{ba} \tau^b_{\,\,c}
+ \Gamo{}^c_{cb} \tau^b_{\,\,a} + 
\left ( \frac{1}{2}  \nn \partial_b \ll + F_{bc} n^c \right ) \tau^{b}_{\,\,a}
+ \tau^{bc} \ell_c U_{ba} = & \\
\frac{1}{\sqrt{|\det (\mathbb{A})|}} \partial_b \left ( \sqrt{| \det (\mathbb{A})|}
\tau^b_{\,\,a} \right )
 - \Gamo{}^c_{ba} \tau^b_{\,\,c} + \tau^{bc} \ell_c U_{ba} = &  [ J_a ].
\end{align*}
To elaborate this further we use the explicit expression for $\Gamo{}^c_{ab}$ in 
Definition \ref{DefGamo} and use the facts $P^{cd} \tau^b_{c} =
P^{cd} \tau^{ba} \gamma_{ac} = \tau^{ba} \left ( \delta^d_a - n^d \ell_a \right ) =
\tau^{db} - n^d \tau^{bc} \ell_c$ 
and $n^c \tau^b_{\,\,c} = n^c \tau^{ba} \gamma_{ac} = - \nn \tau^{bc} \ell_c$. Hence
\begin{align*}
- \Gamo{}^c_{ba} \tau^b_{\,\,c} + \tau^{bc} \ell_c U_{ba}  = & 
- \frac{1}{2} \tau^{bd}  \partial_a \gamma_{dc} 
+ \tau^{bc} \ell_c \left (  
\frac{1}{2} n^d \left ( \partial_b \gamma_{da} + \partial_a \gamma_{db} - \partial_d \gamma_{ba} \right )
+ \frac{1}{2} \nn \left ( \partial_a \ell_b + \partial_b \ell_a \right )
+ U_{ba}  \right ), \\
= & - \frac{1}{2} \tau^{bd}  \partial_a \gamma_{dc} 
\! + \! \frac{1}{2} \tau^{bc} \ell_c \left (  
- \pounds_{\hat{n}} \gamma_{ba}
+ \partial_b \left ( \gamma_{da} n^d \right )
+ \partial_a \left ( \gamma_{ da} n^d \right )
+ \nn \partial_a \ell_b + \nn \partial_b \ell_a + 2 U_{ba}
\right ) \\
= & - \frac{1}{2} \tau^{bd}  \partial_a \gamma_{dc},  
\end{align*}
where in the last equality we have used the definition of $U_{ba}$ (see Lemma \ref{identitiesnablao})
and  $\gamma_{ad} n^d = - \nn \ell_a$. Putting things together, equation (\ref{shellJ}) becomes
\begin{eqnarray*}
\frac{1}{\sqrt{|\det (\mathbb{A})|}} \partial_b \left ( \sqrt{| \det (\mathbb{A})|}
\tau^b_{\,\,a} \right ) - \frac{1}{2} \tau^{bd} \partial_a \gamma_{bd} =  [ J_a ].
\end{eqnarray*}
We can summarize the result in the following Lemma
\begin{lemma}
Let $\{\N,\gamma_{ab},\ell_a,\ll,Y^{\pm}_a\rho_{\ell}^{\pm}, J^{\pm}_{c}\}$
be a shell. In coordinates, the shell field equations take the form
\begin{align*}
\frac{1}{\sqrt{|\det (\mathbb{A})|}} \partial_a \left ( \sqrt{| \det (\mathbb{A})|} 
\tau^{ab} \ell_b \right ) - \frac{1}{2} \tau^{ab} \left ( Y^+_{ab} + Y^-_{ab} \right ) = [ \rho_{\ell} ], \\
\frac{1}{\sqrt{|\det (\mathbb{A})|}} \partial_b \left ( \sqrt{| \det (\mathbb{A})|}
\tau^b_{\,\,a} \right ) - \frac{1}{2} \tau^{bd} \partial_a \gamma_{bd} =  [ J_a ],
\end{align*}
where $\mathbb{A}$ is given in Definition \ref{hypmetdata}. 
\end{lemma}
We  finish this section  with an analysis of the
behaviour of the shell equations under a gauge transformation. To that end let us
define the following tensors
\begin{align*}
B \defi &
\frac{1}{\sqrt{|\det (\mathbb{A})|}} \partial_a \left ( \sqrt{| \det (\mathbb{A})|} 
\tau^{ab} \ell_b \right ) - \frac{1}{2} \tau^{ab} \left ( Y^+_{ab} + Y^-_{ab} \right ) - [ \rho_{\ell} ], \\
C_a \defi &  
\frac{1}{\sqrt{|\det (\mathbb{A})|}} \partial_b \left ( \sqrt{| \det (\mathbb{A})|}
\tau^b_{\,\,a} \right ) - \frac{1}{2} \tau^{bd} \partial_a \gamma_{bd} - [ J_a ].
\end{align*}
So that the shell equations are simply $B =0$ and $C_a =0$. The gauge behaviour of these
fields is as follows
\begin{proposition}
Under a gauge transformation with gauge fields $(u,V^a)$ the tensors $B$, $C_a$ defined above
transform as
\begin{eqnarray}
C'_a = \frac{1}{u} C_a, \quad \quad
B' = B + V^a C_a. \label{invshelleqs}
\end{eqnarray}
\end{proposition}

\vs

\noindent {\bf  Remark \arabic{remark}. \addtocounter{remark}{1}} This gauge behaviour provides a powerful consistency check for the validity of the
equations. Indeed, the validity of equation $B=0$ for an arbitrary gauge implies immediately
the validity of the equation $C_a=0$. So, we could have concentrated on the equation for 
$[\rho_{\ell}]$ alone and derive the equation for $[J_{a}]$ as a consequence of the gauge freedom, which certainly gives strong support to the validity of both
equations.

\vs

\noindent {\it Proof.}  We start finding the gauge behaviour of $\det(\mathbb{A})$, which reads explicitly
\begin{eqnarray}
\det ( \mathbb{A'} ) =
\det 
\begin{bmatrix}
\gamma_{ab} & u \left ( \ell_a + V^c \gamma_{ac} \right ) \\
u \left ( \ell_b + V^d \gamma_{bd} \right )
 & u^2 \left ( \ll + 2 V^c \ell_c + V^d V^d \gamma_{cd} \right )  
\end{bmatrix}.
\label{Aprime}
\end{eqnarray}
Multiplying the column $b$ by $- u V^b$ ($b = 1,2,3$) and adding the 
three columns yields
\begin{eqnarray*}
\begin{bmatrix}
- u \gamma_{ab} V^b \\
 - u^2 \left (\ell_b V^b + V^d \gamma_{bd} V^b \right )
\end{bmatrix}.
\end{eqnarray*}
Adding this 
to the last column in the matrix (\ref{Aprime})
does not change its determinant. Hence
\begin{eqnarray*}
\det ( \mathbb{A'} ) =
\det 
\begin{bmatrix}
\gamma_{ab} & u \ell_a  \\
u \left ( \ell_b + V^d \gamma_{bd} \right )
 & u^2 \left ( \ll +  V^c \ell_c \right )  
\end{bmatrix}.
\end{eqnarray*}
We now repeat the process but now multiplying the row $a$ by $-u V^a$ and  adding them together to give the 
row vector $( - u \gamma_{ab} V^a, - u^2 \ell_a V^a )$. 
Adding this to the fourth row in the matrix does not change the determinant, so
\begin{eqnarray*}
\det ( \mathbb{A'} ) =
\det 
\begin{bmatrix}
\gamma_{ab} & u \ell_a  \\
u \ell_b  & u^2 \ll   
\end{bmatrix} = u^2 \det ( \mathbb{A} ),
\end{eqnarray*}
where the last equality follows, for instance, by multiplying the first three
rows and first three columns by $u$ and extracting a common factor $u^2$ to the matrix.
Recalling the gauge transformation (\ref{break}) for $\tau^{b}_{\,\,a}$ it follows that 
$\sqrt{|\det ( \mathbb{A} )|} \tau^b_{\,\,a}$ is gauge invariant. Since $J^{\prime}_a = 
u^{-1} J_a$ and  $\gamma_{bd}$ is gauge invariant $C'_a = u^{-1} C_a$ follows at once.

In order to prove the second expression in (\ref{invshelleqs}) we recall the transformation laws
$T^{\prime}{}^a = T^a + V^b \tau^{a}_{\,\,b}$, $[\rho'_{\ell}] = 
[\rho_{\ell}] + V^a J_{a}$,  as well as (\ref{gaugetrans}) for $Y_{ab}$ to write
\begin{align*}
B' = &  \frac{1}{u \sqrt{|\det(\mathbb{A})|}} 
\partial_a \left ( u \sqrt{|\det(\mathbb{A})|} \left ( \tau^{ab} \ell_{b} 
+ V^b \tau^a_{\,\,b} \right ) \right ) - \frac{1}{u} \tau^{ab}
\left ( u \overline{Y}_{ab} + \ell_a \partial_b u
 + \frac{1}{2} \pounds_{u \hat{V}} \gamma_{ab}
\right ) - [\rho_{\ell}] - V^a [J_{a}]  \\
%\frac{\partial_a u}{u} \left ( \tau^{ab} \ell_{b} + V^b \tau^{a}_{\,\,\,b} 
%\right )
%+ B + V^b \frac{1}{\sqrt{|\det(\mathbb{A})|}} \partial_a \left ( 
%\sqrt{|\det(\mathbb{A})|} \tau^a_{\,\,b} \right )
%+ \tau^a_{\,\,\,b} \partial_a V^b 
%- \frac{1}{u} \tau^{ab} \ell_b \partial_a u 
%- \frac{1}{2u} \tau^{ab} \pounds_{u \hat{V}} \gamma_{ab} 
%- V^a [J_{a} ] = \\
= & B + V^a C_a +  \frac{1}{u} \tau^a_{\,\,b} \partial_a \left ( u V^b
\right ) + \frac{1}{2} V^a \tau^{bd} \partial_a \gamma_{bd} 
- \frac{1}{2u} \tau^{ab} \pounds_{u \hat{V}} \gamma_{ab}  = B + V^a C_a,
\end{align*}
where in the second equality we have expanded the derivatives and
recalled the definition of $B$ and $C_a$ and 
in the last one we have used the coordinate expression
for $\pounds_{u \hat{V}} \gamma_{ab}$. \ep

\section{Conclusions}
\label{Conclusions}

In this paper we have obtained a consistent framework to define
data on $m$-dimensional manifolds consistent with the geometry of hypersurfaces
of arbitrary causal character in $(m+1)$-dimensional spacetimes. We have also
obtained explicitly the form of the the constraint equations (i.e.
the normal-tangential and normal-transversal components of the Einstein 
tensor in terms of hypersurface data) when the data is embedded in a
spacetime and have thus defined the constraint equations at an abstract
hypersurface level. As a simple application we have derived the shell
equations arising from two hypersurface data which agree except for
its transverse tensor $Y_{ab}$. In a spacetime setting, such data
arises in the matching theory of two spacetimes, hence the name 
shell equations. These equations generalize the well-known shell
equations derived in the matching of spacetimes across spacelike,
timelike or null boundaries. 

In a remarkable paper \cite{JezierskiKijowski2002} (see also
\cite{JezierskiKijowski2000, Jezierski2004}),
the constraint equations for the normal-tangential 
component of the Einstein tensor in the case of null hypersurfaces was
derived in full generality using an interesting geometric property
of null hypersurfaces, namely that there exists an intrinsic (i.e.
coordinate independent) derivative on the null hypersurface capable of
evaluating the divergence of  tensor densities of the form
$H^{a}_{\,\,\,b}$ satisfying $H^a_b n^b =0$ and $\gamma_{ac} H^c_{\,\,b}
= \gamma_{bc} H^{c}_{\,\,\,a}$. This derivation depends only on the degenerate
first fundamental form $\gamma_{ab}$ of $\N$. In the results above we have dealt with
data of arbitrary causal character which obviously, must
agree with the construction in \cite{JezierskiKijowski2002}  in the null case. In fact, as
mentioned in the Introduction, there
exist several distinct approaches to the initial data and constraint equations
in the characteristic case.
The framework above, being completely general, should be
useful in trying to clarify the relationship between the various approaches
to the constraint equations for characteristic initial data. This will be
the subject of a future investigation.

\section*{Acknowledgements}
I wish to thank Jos\'e M.M. Senovilla and Alberto Soria for useful comments
on a previous version of the manuscript.
Financial support under the projects FIS2012-30926 (Spanish MICINN)
and P09-FQM-4496 (Junta de Andaluc\'{\i}a and FEDER funds).

\end{document}